\documentclass[fleqn,usenatbib,preprint,latterpaper]{mnras}

\usepackage[T1]{fontenc}
\usepackage{ae,aecompl}

\usepackage{etoolbox}
\makeatletter
\patchcmd\@combinedblfloats{\box\@outputbox}{\unvbox\@outputbox}{}{%
  \errmessage{\noexpand\@combinedblfloats could not be patched}%
}%
\makeatother

\usepackage{graphicx}	
\usepackage{amsmath}	
\usepackage{amssymb}	
\usepackage[usenames, dvipsnames]{color} 
\usepackage{float}
\usepackage{subcaption}
\usepackage{verbatim}

\title
[UHECRs from shocks in the lobes of radio galaxies]
{Ultra-high energy cosmic rays from shocks in the lobes of
powerful radio galaxies}

\author[J.~H.~Matthews et al.]{J.~H.~Matthews$^1$\thanks{james.matthews@physics.ox.ac.uk}, A.~R.~Bell$^2$, K.~M.~Blundell$^1$
and A.~T.~Araudo$^3$
\\$^1$University of Oxford, Astrophysics, Keble Road, Oxford, OX1 3RH, UK
\\$^2$University of Oxford, Clarendon Laboratory, Parks Road, Oxford OX1 3PU, UK
\\$^3$Astronomical Institute, Academy of Sciences of the Czech Republic,
Bo\v{c}n\'{\i} II 1401, CZ-141\,00 Prague}

\date{Accepted to MNRAS 2018 October 26. Received 2018 October 16; in original form 2018 August 10.}

\pubyear{2018}

\begin{document}
\label{firstpage}
\pagerange{\pageref{firstpage}--\pageref{lastpage}}
\maketitle

\begin{abstract}
The origin of ultra-high energy cosmic rays (UHECRs) has been 
an open question for decades. Here, we use a combination of 
hydrodynamic simulations and general physical arguments 
to demonstrate that UHECRs can in principle be produced by
diffusive shock acceleration (DSA) in shocks in the backflowing 
material of radio galaxy lobes. These shocks occur 
after the jet material has passed through the relativistic 
termination shock. Recently, several authors have demonstrated that 
highly relativistic shocks are not effective in accelerating UHECRs. 
The shocks in our proposed model have a range of non-relativistic or mildly relativistic shock velocities more conducive to UHECR acceleration, with shock sizes in the range $1-10$~kpc. 
Approximately 10\% of the jet's energy flux is focused through
a shock in the backflow of $M>3$. 
Although the shock velocities can be
low enough that acceleration to high energy via DSA is still
efficient, they are also high enough for the Hillas energy to approach 
$10^{19-20}$~eV, particularly for heavier CR composition and 
in cases where fluid elements pass through 
multiple shocks. We discuss some of the more general considerations 
for acceleration of particles to ultra-high energy with reference
to giant-lobed radio galaxies such as Centaurus A and Fornax A, 
a class of sources which may be responsible for the 
observed anisotropies from UHECR observatories. 
\end{abstract}



\def\bb{\boldsymbol}
\def\alfven{Alfv\'en}
\newcommand{\pdiff}[2]{\ensuremath{\frac{\partial #1}{\partial#2}}}
\newcommand{\mydiff}[2]{\ensuremath{\frac{d #1}{d#2}}}
\newcommand{\mydiv}[1]{\ensuremath{\nabla \cdot #1}}
\def\pluto{\textsc{Pluto}}

\section{Introduction}
Ultra-high energy cosmic rays (UHECRs) are 
cosmic rays (CRs) that arrive at Earth
with energies extending beyond $10^{20}$~eV. 
Although the acceleration of Galactic CRs with energies of about $100$~TeV in supernova remnants (SNRs) is well-established 
\citep{volk_magnetic_2005,uchiyama_extremely_2007,bell_particle_2014}, 
as yet, the origin of UHECRs is not known. 
They must be extragalactic, since their Larmor radius in 
a reasonable background magnetic field is larger than 
the Galactic scale height ($\sim 1$~kpc), 
but they must originate from within 
a few mean free paths for attenuation by the Greisen--Zatsepin--Kuzmin \citep[GZK;][]{greisen_end_1966,zatsepin_upper_1966} effect
and photodisintegration \citep[e.g.][]{stecker_photodisintegration_1999}.
Both processes have a typical attenuation length of $\sim50-100$Mpc 
\citep[e.g.][]{alves_batista_crpropa_2016,wykes_uhecr_2017}. 
Furthermore, any complete 
production model must explain the observed anisotropies 
\citep{abu-zayyad_search_2013,abbasi_indications_2014,pierre_auger_collaboration_observation_2017,pierre_auger_collaboration_indication_2018} 
and the composition of CRs at high energies \citep{pierre_auger_collaboration_depth_2014,de_souza_measurements_2017}.
Meeting all these requirements simultaneously is a challenge. 

One of the best candidate 
mechanisms for accelerating CRs to high energy 
is diffusive shock acceleration 
\citep[DSA;][]{axford_acceleration_1977,krymskii_regular_1977,blandford_particle_1978,bell_acceleration_1978,bell_acceleration_1978-1},
also known as first order Fermi acceleration. 
The characteristic maximum energy a CR can gain by this process is set by the 
Hillas energy \citep{hillas_origin_1984}, given by
\begin{equation}\label{E-Hillas}
E_H = 0.9~\mathrm{EeV}~Z
\left( \frac{B}{\mu \mathrm{G}} \right)
\left( \frac{v_s}{c} \right)
\left( \frac{r}{\mathrm{kpc}} \right),
\end{equation}
where $B$ is the magnetic field, $v_s$ is the shock velocity,
$Z$ is the atomic number of the nucleus
and $r$ is the characteristic size of the shock. 
While other mechanisms such as {\em second-order} Fermi 
acceleration \citep{fermi_origin_1949}, shock drift 
acceleration 
\citep[e.g.][]{armstrong_shock_1985,burgess_shock_1987,decker_role_1988}
and ``one-shot'' mechanisms 
\citep[e.g.][]{litvinenko_particle_1996,haswell_high-energy_1992,caprioli_espresso_2015}
may also work, DSA is attractive 
since it naturally produces a power law similar to that observed
and also probably accelerates the electrons in SNR and radio 
galaxies to high energy where radiation is clearly 
emitted near shock fronts \citep[e.g.][]{laing_radio_1989,koyama_evidence_1995}. 

Given the dependence of the Hillas energy on shock size
and speed, it is natural to turn to the largest systems we know 
of that show energetic outflows and strong shocks. 
In this sense, active galactic nuclei (AGN) and their associated outflows
make for obvious UHECR candidate sources \citep[e.g.][]{hillas_origin_1984, norman_origin_1995,osullivan2009,hardcastle_which_2010}. 
AGN can launch dramatic jets from close to the black 
hole, which can then travel for hundreds of kpc into the surrounding 
medium, producing giant radio galaxies. 
Radio galaxies typically fall into one of two \cite{fanaroff_morphology_1974}
classes; class I sources (FRIs) are brightest at the centre and 
have fairly low power jets that entrain material, becoming 
disrupted relatively close to the nucleus, whereas 
class IIs (FRIIs) can proceed uninterrupted for long 
distances, showing bright `hotspots' at the ends of the 
jets where they produce a termination shock. 

Catalogues of AGN and radio galaxy positions 
can be correlated with the arrival directions from
the Pierre Auger Observatory (PAO) and other UHECR
detectors. Initial PAO results suggested an
association with AGN \citep{pierre_auger_collaboration_correlation_2007,pierre_auger_collaboration_correlation_2008}, but updated results
were of lower significance \citep{abreu_update_2010}. 
However, more recently significant departures from
isotropy have been observed 
\citep{pierre_auger_collaboration_observation_2017,pierre_auger_collaboration_indication_2018}, with 
\cite{pierre_auger_collaboration_indication_2018} finding significant correlations 
of 4 and 2.7$\sigma$ with starburst galaxies and AGN, respectively.
However, as we showed in a recent 
Letter \citep{matthews_fornax_2018_mnras}, considering the most 
recent {\sl Fermi} catalogues and accounting for magnetic 
deflection can increase the correlation with AGN, 
particularly if the local contribution to UHECRs is dominated
by Fornax~A and Centaurus~A. 

Despite the promise of radio galaxies as UHECR sources, 
relativistic shocks such as their termination shocks
are actually rather poor accelerators of UHECRs 
\citep{lemoine_electromagnetic_2010,kirk_radiative_2010,reville_maximum_2014,bell_cosmic-ray_2018}.
In a recent paper, we showed that the maximum energy in an 
ultra-relativistic shock is well below the EeV range 
\citep{bell_cosmic-ray_2018}. We also applied similar
arguments to the observed radio spectra in the hotspots of 
Cygnus A \citep{araudo_maximum_2018} and other FRII sources
\citep{araudo_particle_2015,araudo_evidence_2016}.
These studies show that while magnetic field 
amplification to above $100\mu$G can occur, the maximum 
energy of the non-thermal electrons is rather low, on 
the order of a TeV. This maximum energy and 
associated synchrotron cutoff is set by the detailed plasma 
physics and ability to drive Larmor-scale turbulence
at the shock, rather than synchrotron cooling, and thus 
this limit also applies to CR protons and nuclei; the limit
can however be relaxed if there is pre-existing turbulent 
magnetic field on the right scale to scatter particles.
Nonetheless, it seems that if radio galaxies are to accelerate 
UHECRs via DSA then a balance must be struck between allowing the Hillas energy 
to be high enough, and not inhibiting the self-regulating 
acceleration process. In other words, a ``goldilocks'' zone in which
the shock parameters are ``just right'' for efficient DSA to 
high energy must exist. The motivation for this paper is to search for 
shocks in radio galaxies that meet these requirements 
(i.e. not the relativistic termination shock) and 
offer favourable conditions for CR acceleration to ultra-high energy.

The paper is structured as follows. We introduce our numerical
method in section 2, before describing the simulation results in section 3.
In section 4 we use some simple Bernoulli-like arguments to study the flow of
plasma in the jet lobe and cocoon. In section 5, we use a combination of 
Lagrangian and Eulerian methods to calculate the shock properties in the 
simulation, which are then used to estimate the maximum CR energy in section 6. 
We discuss our results in section 7, with particular reference to radio galaxy 
luminosity functions, power requirements and results from UHECR observatories, 
before concluding in section 8.
We adopt the convention of referring to a {\em cocoon} as all the shocked 
jet material that enshrouds the jet beam, and a {\em lobe} as the cocoon
material close to the hotspot that is typically observed in radio. 
We refer to kinetic powers, radiative luminosities and pressures with the
symbols $Q$, $L$ and $P$, respectively. 

\section{Numerical Method}
We use the freely available Godunov-type Eulerian
code \pluto\ \citep{mignone_pluto:_2007} 
to solve the equations of relativistic hydrodynamics 
(RHD), which can be written as
\begin{align}
\frac{\partial D}{\partial t} &=
- \nabla \cdot ( D \boldsymbol{v}), \\
\frac{\partial \bb{m}}{\partial t} &=
- \nabla \cdot ( \bb{m} \boldsymbol{v}) - \nabla P, \\
\frac{\partial E}{\partial t} &=
-\nabla \cdot \boldsymbol{m}. 
\end{align}
Here, $\bb{v}$ is the three-velocity and $P$ is the 
pressure. The conserved quantities are 
$D=\rho \Gamma $, $\bb{m} = \rho h \Gamma^2 
\bb{v}$ and $E = \rho h \Gamma^2 - P$, where $\rho$, 
$\Gamma$ and $h=1+e+P/\rho$ are the density, Lorentz factor 
and specific enthalpy, respectively.

We adopt the Taub-Matthews equation of state 
\citep{mignone_piecewise_2005}.
We use a dimensionally unsplit scheme with 
second-order Runge-Kutta time integration and a Courant-Friedrichs-Levy number
of 0.4 in 2D and 0.3 in 3D. We employ the
monotonized central (MC) limiter on characteristic variables, the HLLC solver and a multi-dimensional shock flattening algorithm. Shock flattening and detection is discussed further in section~\ref{sec:shock_detect}. We also inject a standard passive scalar jet tracer, $C_j$, which is advected according to the equation
\begin{equation}
\frac{\partial (\rho C_j)}{\partial t} = - \nabla \cdot (\rho C_j 
\bb{v}).
\end{equation}

\subsection{Jet and Cluster Setup}
We initially set up the background cluster 
density profile with an isothermal  
$\beta$ profile 
or King profile \citep{king_density_1972}, given by 
\begin{equation}
\rho(r) = \rho_0 \left[ 1 + 
\left(\frac{r}{r_c} \right)^2
\right]^{-3\beta/2}
\end{equation}
where $r$ is the distance from the centre of the cluster, 
the exponent $\beta$ is an input parameter, 
$\rho_0$ is the density at $r=0$ 
(in this case equal to the simulation unit density)
and $r_c$ is the scale length or core radius. 
We set $r_c=50$kpc and $\beta=0.5$ to
roughly match the median values from \cite{ineson_link_2015}
for a sample of clusters hosting 
radio-loud AGN. We have also verified that
the absolute pressure and density values are within the 
range of those observed. The pressure in the cluster is set 
so that the sound speed is a constant value of 
$515.8$~km~s$^{-1}$. This corresponds to a temperature 
of about 1~keV, typical for radio galaxy environments 
\citep{ineson_radio-loud_2013}. We impose a gravitational 
potential derived from the pressure gradient 
so that the cluster atmosphere is initially in
hydrostatic equilibrium. In the 3D simulation we also
apply small random density perturbations 
($\delta \rho / \rho \approx 10^{-10}$) 
in the environment so there is an imposed departure from 
rotational symmetry. 

The jet is injected via a nozzle of radius $r_j$ at the origin (in 2D cylindrical),
or at the centre of the $x-y$ plane (in 3D cartesian), and the inflow boundary is
smoothed with the otherwise reflective boundary condition 
at $z=0$ using a $1/\cosh(r)$ profile. 
For a rest mass density contrast of $\eta=\rho_j/\rho_0$,
where $\rho_j$ is the jet density,
the relativistic generalisation of the jet density ratio is given by 
\citep[e.g.][]{marti_morphology_1997,krause_very_2005}
\begin{equation}\label{eta-rel}
\eta_r = \frac{\rho_j h_j}{\rho_0 h_0} \Gamma_j^2,
\end{equation}
where $\Gamma_j$ is the Lorentz factor of the jet and 
$h_j$ and $h_0$ are the jet and environment
specific enthalpies, respectively. Low $\eta_r$ 
corresponds to a high density contrast between the jet and 
ambient medium. The jet power for a top-hat jet is given by
\begin{equation}
Q_j = \pi r_j^2 v_j \left[ \Gamma_j (\Gamma_j-1) \rho_j c^2 
+ \frac{\gamma}{\gamma - 1} \Gamma^2 P_j
\right].
\label{eq:jet_power}
\end{equation}
The properties of the jets and simulation domains for each of our simulations
are listed in Table~\ref{tab:sim_properties}. The actual jet powers
are slightly lower than from equation~\ref{eq:jet_power} due to
the smoothing function applied to the boundary condition; the smoothed values
are given in the table.

\begin{table*}
\centering
    \begin{tabular}{l c c c c c c c c c }
    \hline Run & Dim. & $v_j/c$ & $\Gamma_j$ & $\eta$ & $\eta_r$ & 
    $r_j$ (kpc) & $Q_j$ (erg~s$^{-1}$)
    & Domain size (kpc) & Resolution ($\Delta x$) \\
    \hline 
    S1 & 2D & 0.5 & 1.15 & $7.52\times10^{-5}$ & $10^{-4}$ & 2 & $1.21\times10^{44}$ &  300 x 120 & 0.2kpc \\
    F1 & 2D & 0.95 & 3.20 & $9.71\times10^{-5}$ & $10^{-3}$ & 1 & $2.69\times10^{45}$ &  300 x 120 & 0.2kpc \\
    F3D & 3D & 0.95 & 3.20 & $1.88\times10^{-5}$ & $1.92\times10^{-4}$ & 2 & $1.00\times10^{45}$ & 240 x 120 x 120 &  0.4kpc \\
    \hline 
    \end{tabular}
  \caption{Jet properties for each of our simulations.
  Each simulation is conducted in the same cluster environment with $\beta=0.5$ and $r_c=50$~kpc.}
  \label{tab:sim_properties}
\end{table*}

Our simulations use fairly typical techniques and input parameters for the simulation of relativistic jets being injected into a smooth cluster medium, allowing them to be compared to a number of other numerical studies \citep[e.g.][]{norman_structure_1982,saxton_complex_2002,krause_very_2005,hardcastle_numerical_2013,hardcastle_numerical_2014,english_numerical_2016} as well as analytic and semi-analytic approaches \citep[e.g.][]{falle_self-similar_1991,kaiser_self-similar_1997}. Our jet powers are within the range of those inferred from observations \citep[e.g.][]{blundell_nature_1999,ineson_representative_2017}. The overall energetics and physical conditions in our jets can be considered a reasonable approximation to reality for moderately powerful FRII jets. 

\subsection{Lagrangian Tracer Particles}
In order to track the history of a fluid element in 
detail, we inject a series of tracer particles in the 
jet. These particles are injected in the 2D simulations 
at regular intervals at the jet aperture by generating 
a random number between 0 and $r_j$,
corresponding to the radial distance from the $z$ axis.
The particles are then advected with the local fluid 
velocity, which is obtained from a bilinear interpolation
on the Eulerian grid. The local primitive variables, 
jet tracer value, and velocity divergence are recorded as the 
tracer particle moves through the simulation domain. 
The detailed fluid histories 
provided by the tracer particles are used to 
analyse the bulk flow and obtain shock properties 
(see section~\ref{sec:shock_properties}).

\subsection{Shock Identification}
\label{sec:shock_detect}
To identify shocks in our simulation, we adopt
a similar method to that described by 
\cite{yang_how_2016} and the method already used 
to flag shock zones in \pluto\ \citep{mignone_pluto:_2007}.
Shock zones are flagged if they:
\begin{enumerate}
\item show compression, $\nabla \cdot \bb{v} < 0$; and
\item show a pressure jump, $\Delta P / P_1 > \epsilon_p$, 
\end{enumerate}
where $\epsilon_p$ is an imposed threshold.
The pressure jump across a shock is related to the upstream 
Mach number from the Rankine-Hugoniot conditions \citep[e.g.][]{yang_how_2016}
\begin{equation}
\frac{\Delta P}{P_1} = \frac{2\gamma}{\gamma+1} (M_s^2 -1),
\end{equation}
where $\Delta P=P_2 - P_1$ is the difference of the 
downstream and upstream pressures 
and $M_s$ is the upstream Mach number.
We use both Eulerian (grid-based) and Lagrangian (tracer particle) methods to analyse shocks in our simulations, as described further in section~\ref{sec:shock_properties}.
In the Eulerian method, the pressure jump is computed 
using a three-point undivided difference operator in each cell 
($\tilde{\nabla}P/P$), while when using the Lagrangian tracer particles the jump 
is computed using the equation 
\begin{equation}
\left( \frac{\Delta P}{P} \right)_{L} = 
\frac{P_{t+\Delta t} - P_{t}}{r_{t+\Delta t} - r_{t}} 
\Delta x \,
\end{equation}
where $t$ is the simulation time, $\Delta t$
is the time resolution at which the properties of the Lagrangian particles are recorded 
and $r_t$ is the distance travelled by the particle at time $t$.
This equation ensures that the pressure jumps computed with both the Eulerian
and Lagrangian methods are calculated over the grid scale, $\Delta x$. 
For the purposes of the shock-flattening algorithm,
we adopt $\epsilon_p=10$ to smooth out 
shocks. We use $\epsilon_p=0.2$ for our shock analytics
but also require that the `upstream' Mach number is 
greater than 1. We have verified that this 
shock-detection algorithm does a good job of locating 
the termination shock and reconfinement shocks 
associated with the jet. We identify the upstream region 
in the shock zone as the coordinate
immediately prior to the flagged shock zone. 
We then calculate the shock velocity by 
assuming that the shocks are relatively steady and so the 
shock velocity is approximately equal to the upstream velocity, 
that is $v_s \approx v_1$, and we calculate the Mach 
number of the shock $M_s$ as 
\begin{equation}
M_s = \sqrt \frac{\rho_1 v_1^2}{\gamma P_1}
\end{equation}
Further details are provided in
section~\ref{sec:shock_properties}.

\section{Results}
\label{sec:results}
We conducted a number of simulations but we
focus mainly on three fiducial runs:
S1, a $0.5c$ jet in 2D, F1, a relativistic 
$0.95c$ jet in 2D, and F3D, a 
relativistic $0.95c$ jet in 3D. 
Each jet is injected into an ambient medium with the same density and 
pressure profile. The properties of each run are shown in 
Table~\ref{tab:sim_properties}
and the parameter sensitivity is briefly explored in 
section~\ref{sec:param_sensitivity}.

\begin{figure*}
\centering
\begin{subfigure}{.45\textwidth}
\centering
   \includegraphics[width=\linewidth]{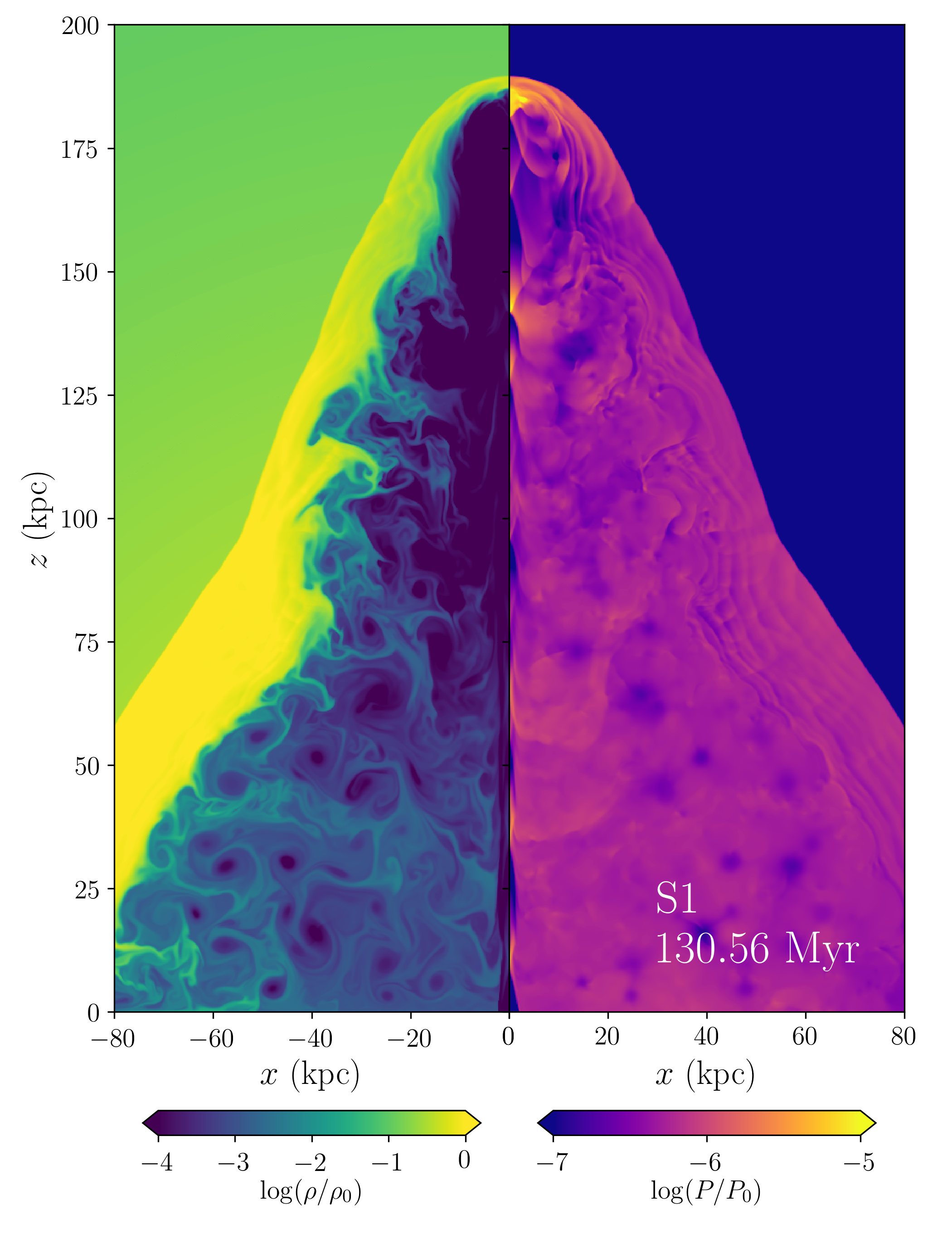}
\end{subfigure}
\begin{subfigure}{.45\textwidth}
\centering
   \includegraphics[width=\linewidth]{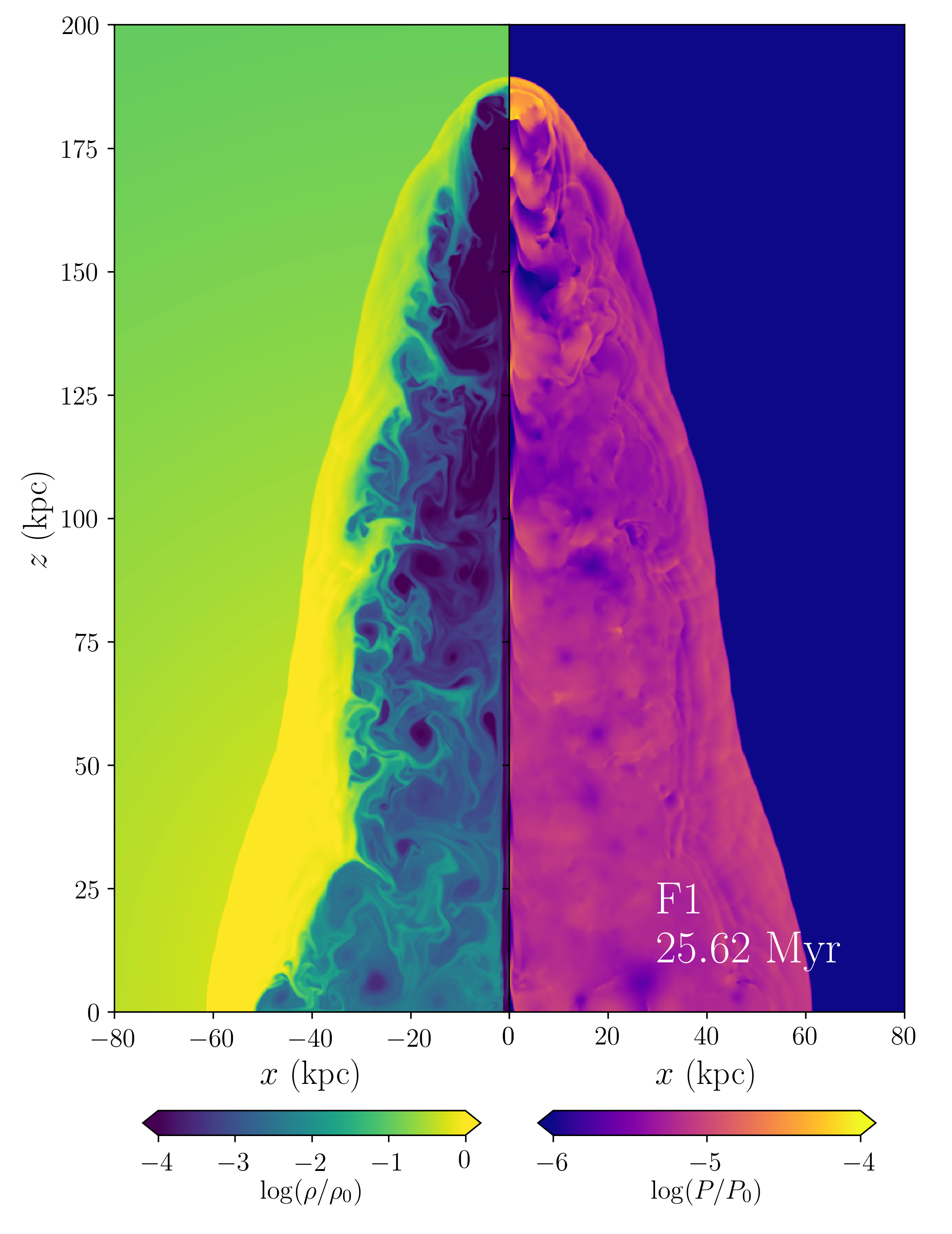}
\end{subfigure}
\caption{Logarithm of density and pressure for a
snapshot of the 2D simulations, S1 (left) and F1 (right). The plots are normalised to the simulation unit density ($\rho_0=6\times10^{-27}$~g~cm$^{-3}$)
and pressure ($P_0=5.393\times10^{-6}$~dyne~cm$^{-2}$).
The jet creates a low-density cocoon which is separated from
the shocked ambient medium by a contact discontinuity.
Sharp shock structures can be seen in the pressure plot.
}
\label{fig:p_and_rho}
\end{figure*}

\begin{figure*}
\centering
\begin{subfigure}{.45\textwidth}
\centering
   \includegraphics[width=\linewidth]{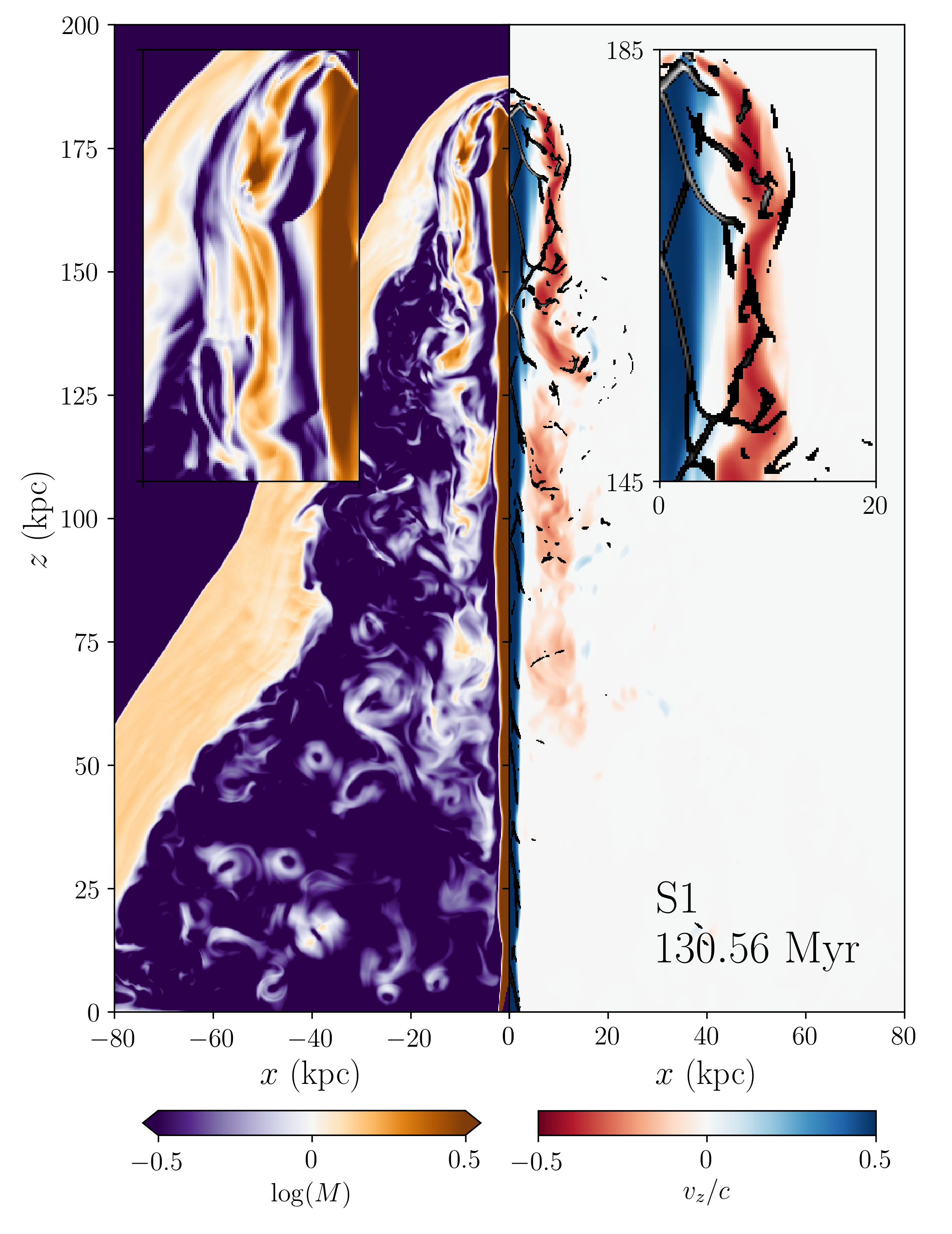}
\end{subfigure}
\begin{subfigure}{.45\textwidth}
\centering
   \includegraphics[width=\linewidth]{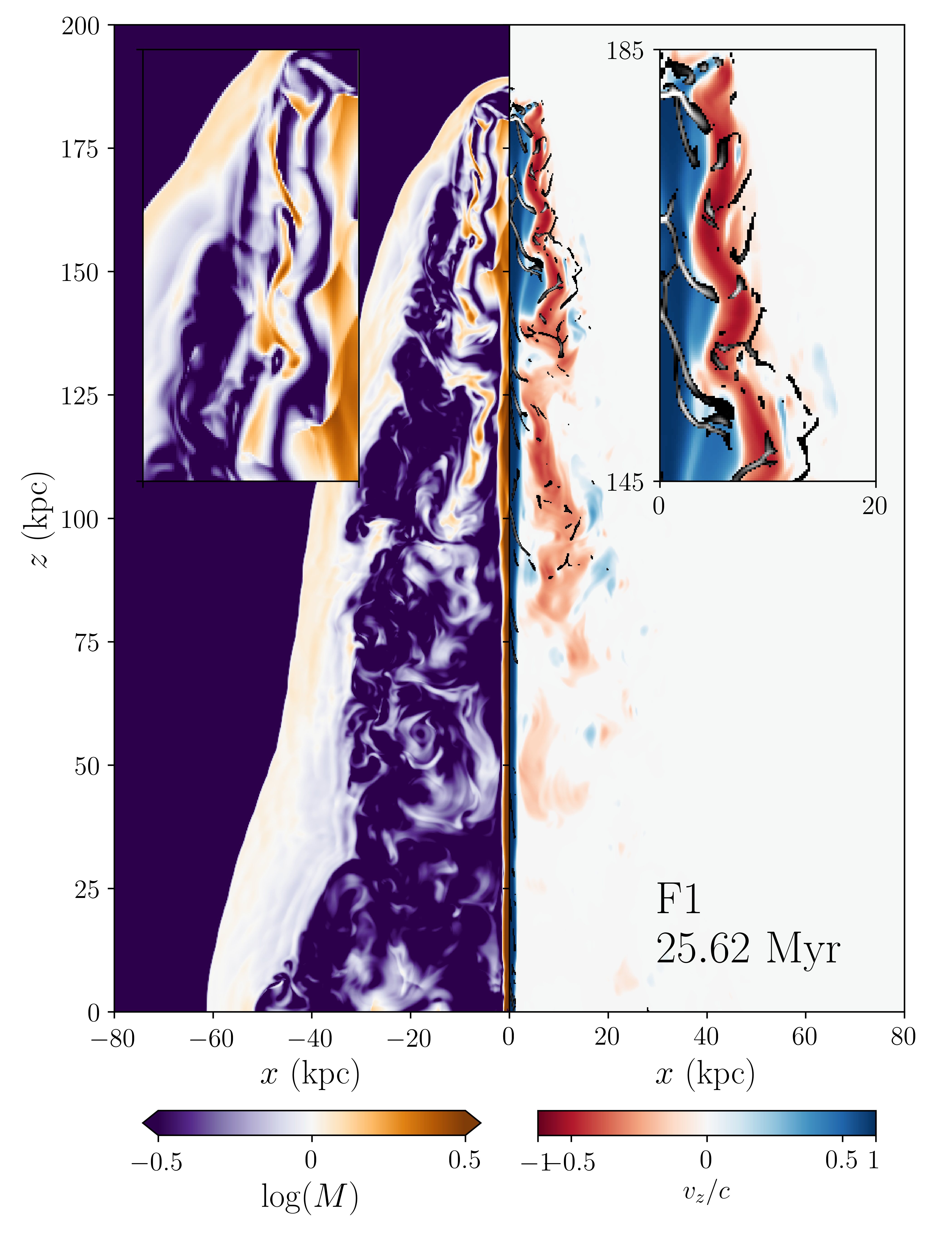}
\end{subfigure}
\caption{Logarithm of Mach number ($M$) 
and vertical velocity ($v_z$) for a snapshot of the 2D 
simulations, S1 (left) and F1 (right). In the $v_z$ plot,
compression structures ($\nabla \cdot \boldsymbol{v} < 0$) 
are coloured in grey to indicate shocks. Supersonic
backflows form in both simulations and vortex shedding
occurs from the jet head. Movies equivalent to these plots
are included in the supplementary material as 
movS1Machvz.mp4 and movF1Machvz.mp4.
}
\label{fig:m_and_dv}
\end{figure*}

\subsection{2D Simulations}
\label{sec:2d_results}
Snapshots of the density and pressure for each of the 
2D simulations are shown in Fig.~\ref{fig:p_and_rho}. In each
of the 2D plots we pick a time stamp at which the jet has travelled
approximately $180$~kpc.The colourmaps are plotted logarithmically and 
normalised to the simulation unit density ($\rho_0=6\times10^{-27}$g~cm$^{-3}$)
and pressure ($P_0=5.393\times10^{-6}$dyne~cm$^{-2}$).
We show the S1 simulation at $130.56$~Myr
and the F1 simulation at $25.62$~Myr.
The jet is launched at high speed from
the base of the grid ($z=0$)
and encounters a series of reconfinement 
shocks as it propagates in the $z$ direction. 
This leads to the Mach number inside the jet dropping from
its initial high value of $10^{10}$ to below $10$, although
it still maintains a high speed close to its initial launch 
velocity. The jet deposits its mechanical energy at a 
termination shock, forming a hot spot. The jet material then 
inflates a low density cocoon. The cocoon and hotspot are 
significantly overpressured with respect to the surroundings and so 
the classic `double-shock' structure is formed, with a 
bow shock propagating into the surrounding ambient medium.
The shocked ambient material is separated
from the shocked jet material by a contact 
discontinuity (CD), although mixing at this CD occurs via the 
Kelvin-Helmholtz instability as well as numerical viscosity on 
the grid scale ($0.2$~kpc in the 2D runs).

In Fig.~\ref{fig:m_and_dv} we show the Mach number, $M$, and 
$z$ (vertical) component of the velocity, $v_z$. In the 
velocity plot we also highlight compression structures in the flow;
pixels are set to a greyscale if, in simulation units of $(c/\mathrm{kpc})$, 
$\nabla \cdot \bb{v}<-0.05$ (F1) or $\nabla \cdot \bb{v}<-0.02$ (S1),
but otherwise are transparent so that the underlying velocity field can 
be seen. The compression structures help highlight the shocks in the 
simulation, which can also be seen to a lesser extent in the pressure plot in
Fig.~\ref{fig:p_and_rho}. Oblique reconfinement shocks can be seen 
clearly up the length of the jet, as well as a clear termination shock
at the jet head. Although the flow is initially subsonic
after the termination shock, it is funnelled sideways 
and backwards, where it becomes supersonic again, producing 
a number of moderately strong shocks. Although Fig.~\ref{fig:m_and_dv} 
is for a single snapshot, 
the movies in the supplementary material along with 
the 3D volume renderings (section~\ref{sec:3d_results}) and  
tracer particle analysis (section~\ref{sec:lagrangian}) make
clear that both the supersonic backflow and associated shock structures 
persist throughout the jet's evolution. 

\subsection{3D Simulations}
\label{sec:3d_results}
\begin{figure*}
\centering
\includegraphics[width=1.0\linewidth, clip=True, trim=0in 0in 0.1in 0in]{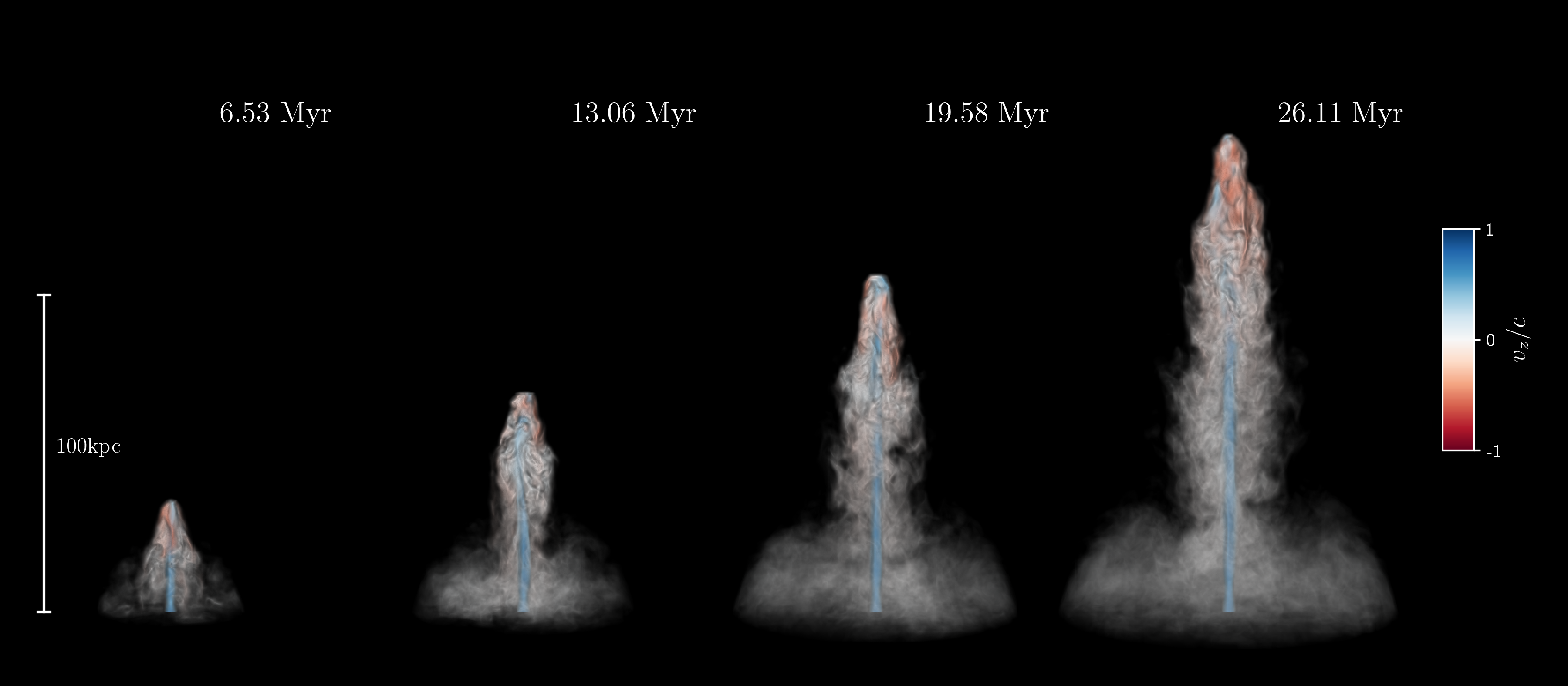}
\caption{Volume rendering of the fiducial fast 3D simulation, F3D,
showing $v_z$, the vertical component of velocity 
at four different times (labelled). The opacity is set linearly
by the jet tracer, $C_j$. A movie equivalent to this plot
is included in the supplementary material as
movF3Dvz.mp4.
}
\label{fig:3d_v}
\end{figure*}

\begin{figure*}
\centering
\includegraphics[width=1\linewidth, clip=True, trim=0in 0in 0.1in 0in]{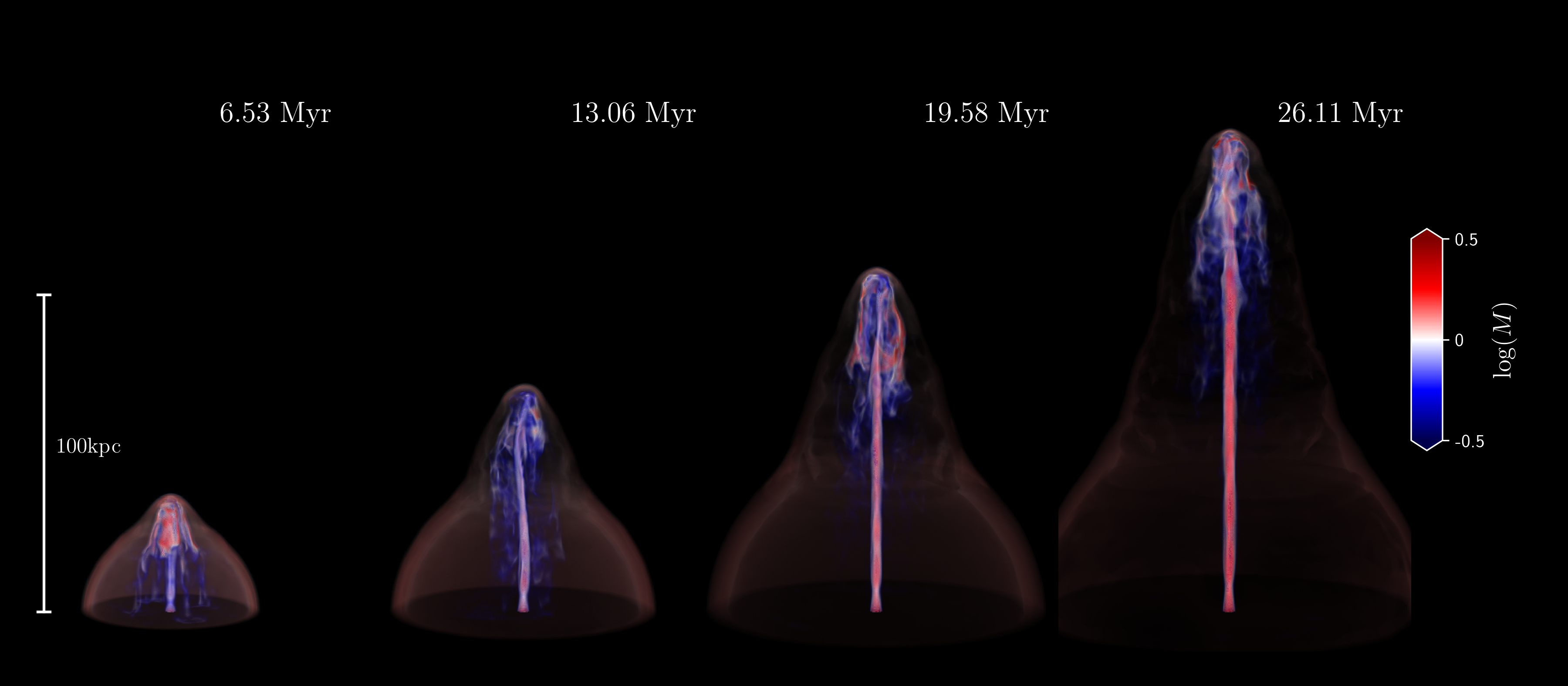}
\caption{Volume rendering of the fiducial fast 3D simulation, F3D,
showing $\log(M)$, the logarithm of the Mach number. Supersonic
flow is coloured red. The opacity is set linearly
by the kinetic energy flux, $1/2 \rho v^3$,
so that the areas in which the kinetic energy is 
focussed can be seen most clearly. A movie equivalent to this plot
is included in the supplementary material as
movF3DMach.mp4.
}
\label{fig:3d_mach}
\end{figure*}

\begin{figure}
\centering
\includegraphics[width=1\linewidth]{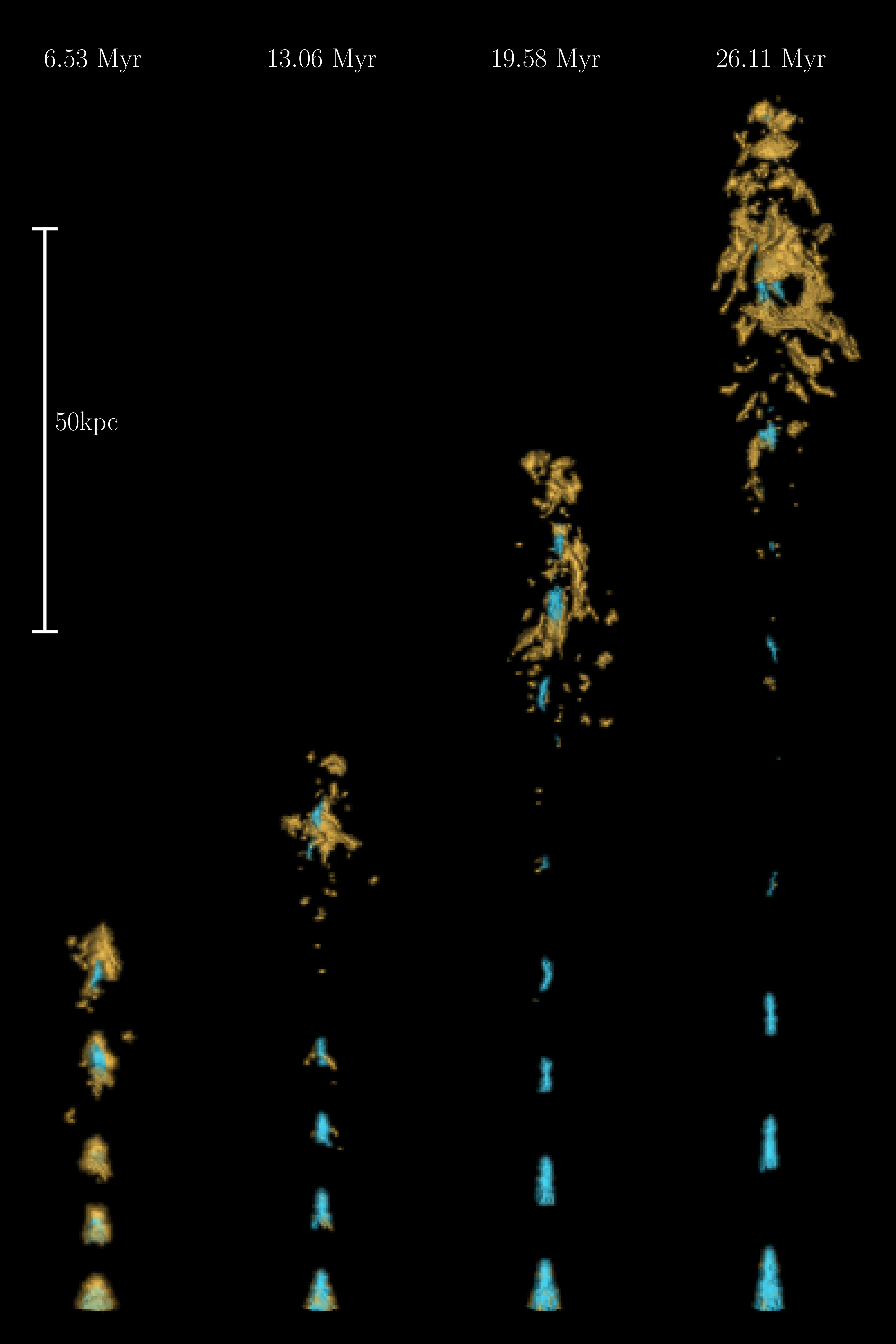}
\caption{Volume rendering of the fiducial fast 3D simulation, F3D,
showing shock regions. Regions are transparent against a black
background unless they satisfy $\tilde{\nabla}{P}/P>0.2$ and $\nabla \cdot \boldsymbol{v} < -0.05 (c/\mathrm{kpc})$, 
in which case they have an opacity of 1. 
Shocks are coloured cyan if they lie within the jet, and
orange if they lie within the lobe or cocoon. 
The reconfinement shocks along the jet axes can be clearly
seen, and there are a number of additional shocks in the lobe region,
similarly to the 2D simulations (see Fig.~\ref{fig:m_and_dv}).}
\label{fig:3d_shocks}
\end{figure}

The 3D simulation (F3D) shows similar behaviour to the 2D simulations,
but with some notable differences. To visualise the simulations, we show volume renderings of $v_z$ and $\log M$ in Figs.~\ref{fig:3d_v} and \ref{fig:3d_mach}. The volume renderings are produced using composite ray-casting in Visit \citep{childs_visit_2005}. In Fig.~\ref{fig:3d_v}, the opacity is set linearly by $C_j$, whereas in Fig.~\ref{fig:3d_mach} the opacity is set linearly by the kinetic energy flux, $\rho v^3/2$. We also show a visualisation of shock structures in Fig~\ref{fig:3d_shocks}, where cells are opaque if they have a pressure gradient of $\tilde{\nabla}{P}/P>0.2$ and $\nabla \cdot \boldsymbol{v} < -0.05 (c/\mathrm{kpc})$. The colour-coding discriminates between shocks in the jet (cyan) and shocks in the lobe or cocoon (orange). All our volume renderings are shown at four different times so the time evolution of the jet can be seen.

The 3D jets propagate more or less uninterrupted to the jet head, where they terminate in a similar manner to the 2D runs. The density perturbations ensure that the otherwise expected n=4 rotational symmetry is broken, leading to a complex flow structure  in the lobe. Fast, supersonic backflows form. These backflowing streams mirror the 2D results in that their velocities can be a significant fraction (up to about half) of the jet velocity, and they persist when the jet has travelled far from the reflective/inflow boundary at $z=0$. However, the backflows can form streams of a helical shape, breaking cylindrical symmetry. This is important, as it means that the energy flux is focused into a smaller cross-sectional area, which can enhance the strength of any shock structures that form. The fraction of the jet power passing through the shock is important in determining the maximum CR energy (see section~\ref{sec:cr_power}). 

\begin{figure*}
\centering
\includegraphics[width=0.8\linewidth]{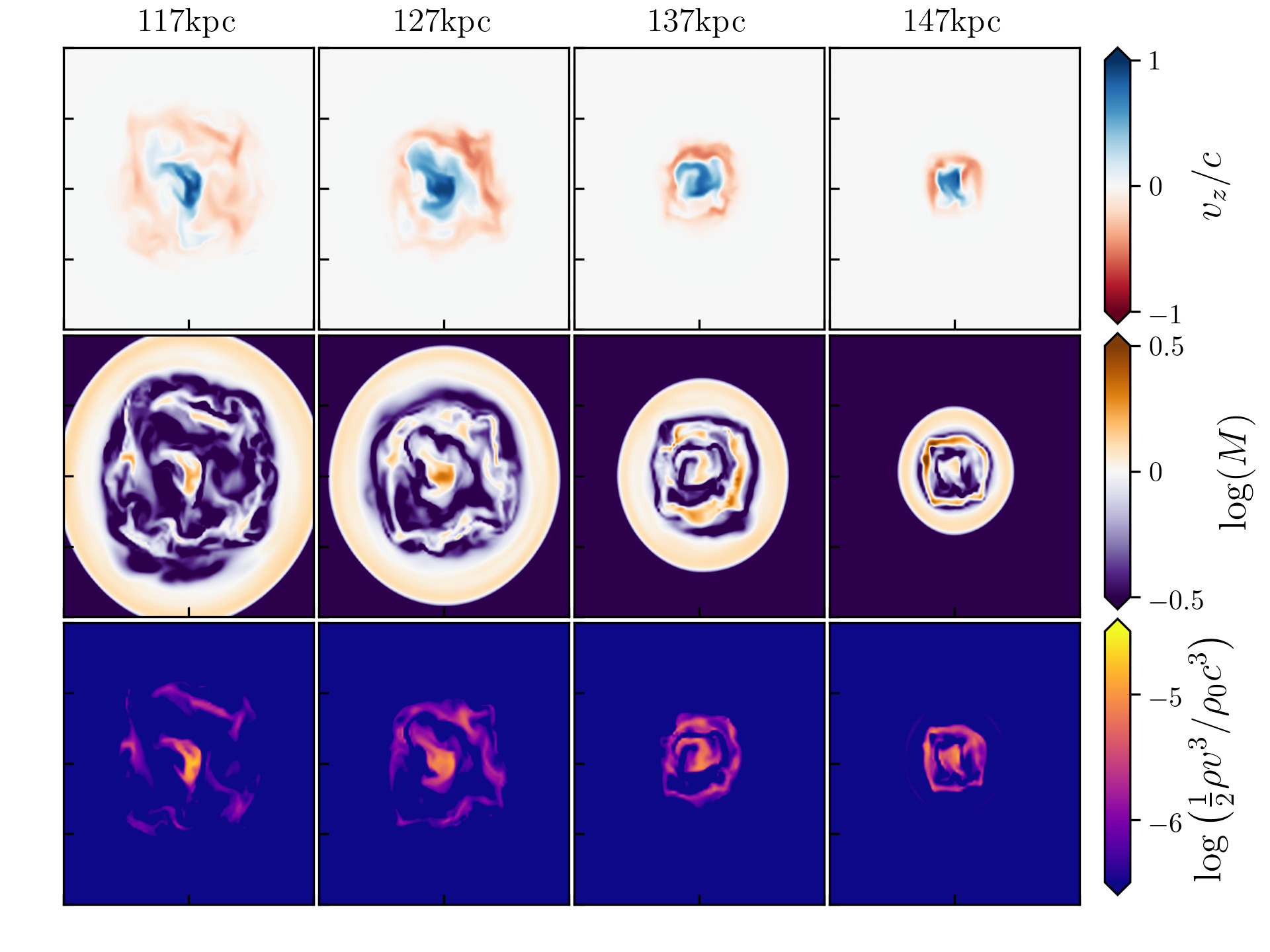}
\caption{Slices in the $x-y$ plane of $z$ velocity component, 
Mach number and kinetic energy flux for the F3D simulation at a timestamp of $26.11$~Myr. 
The slices are $40\mathrm{kpc}\times40\mathrm{kpc}$ and are taken at $10$~kpc intervals when the jet bow shock has advanced approximately $150$~kpc from the injection point. The kinetic energy flux is plotted in simulation units ($\rho_0 c^3$).}
\label{fig:slices}
\end{figure*}

We can gain more insight into the geometry and strength of the backflow in 3D by taking slices in the $x-y$ plane at a few different values of $z$. Slices of $v_z$, $\log(M)$ and kinetic energy flux, $\rho v^3/2$, are shown in Fig.~\ref{fig:slices} for the F3D run at a time stamp of $26.11$~Myr. The plots show cylindrical symmetry to an extent, but a degree of focusing into asymmetric streams occurs. In reality, the exact degree of asymmetry may depend on the environment of the jet and the variation in launch direction, which highlights the importance of simulations that take into account more realistic cluster `weather' \citep[e.g.][]{mendygral_mhd_2012}. These plots make clear that, although a simplification, the 2D simulations in cylindrical symmetry offer a fairly good approximation to the 3D physics, so we just focus on the simpler 2D simulations for our Lagrangian shock analysis, particular as the focusing of the streams suggests that the cylindrically symmetric approximation is conservative if anything. However, a more detailed analysis of 3D simulations is potentially important and should be investigated. We do however analyse the shock sizes in our 3D simulation. 

\begin{figure*}
\centering
\includegraphics[width=0.85\linewidth]{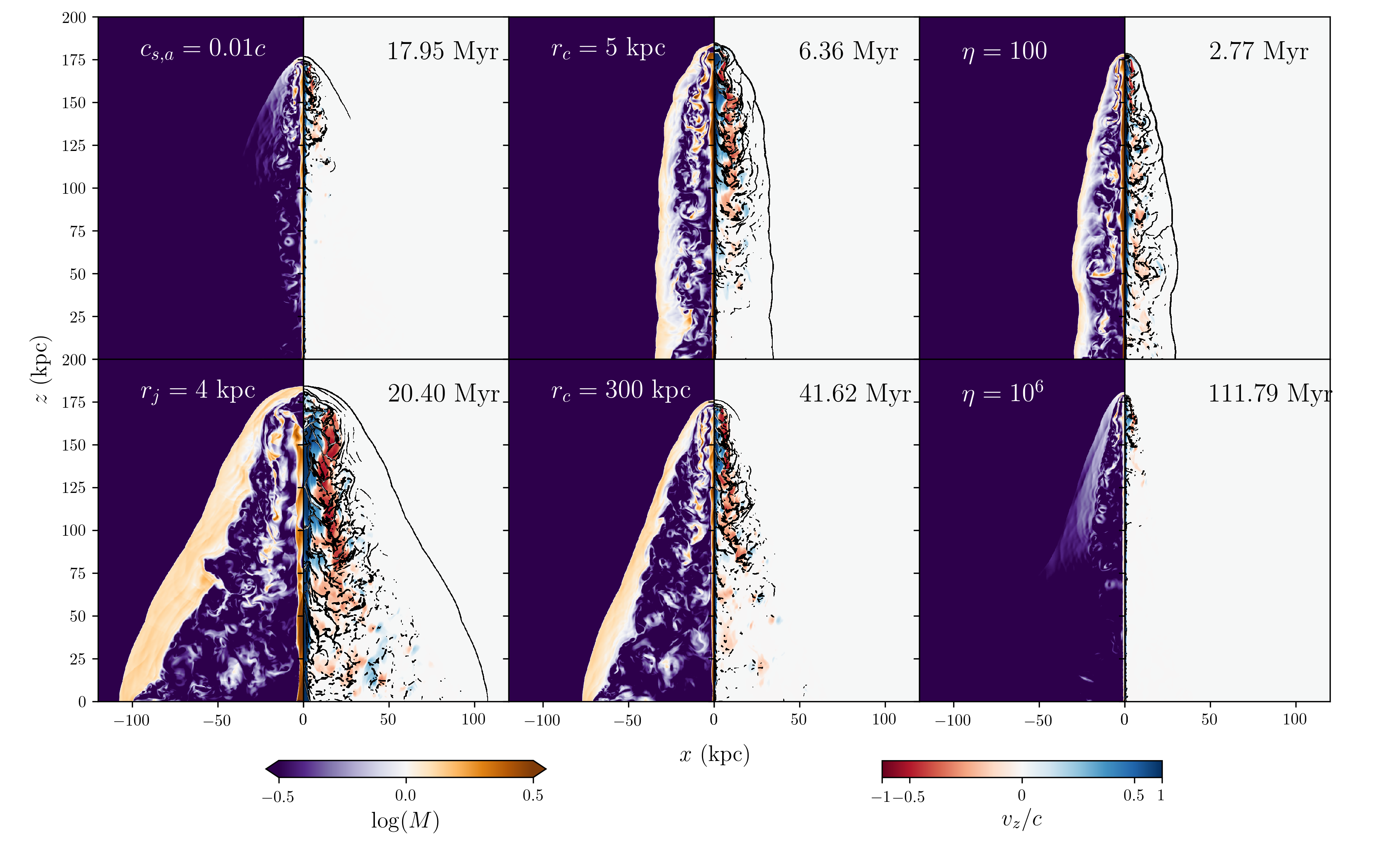}
\caption{The effect of varying input parameters
on the Mach number of the backflow. Analogues to Fig.~\ref{fig:m_and_dv} are shown for 6 separate
runs where one parameter is varied with respect to the F1 simulation. The parameter in question
and the simulation time is marked in each panel. 
Advance speeds and morphologies are varied, but
while the details of the backflows change, they
remain at least partially fast and supersonic 
in all cases.
}
\label{fig:parameters}
\end{figure*}

\subsection{Parameter sensitivity}
\label{sec:param_sensitivity}
It is important to know if the formation of fast, supersonic backflows 
is limited to our chosen region of parameter space. 
A comprehensive exploration of parameter 
space and the resultant backflow dynamics would make for  
interesting future work, but this is beyond the scope of the current
study. However, to briefly highlight the impact of some important
parameters on the jet dynamics and morphology, we explore the effect
of varying $\eta_r$, $c_{s,a}$, $r_c$ and $r_j$. 
Taking the F1 simulation as our starting point, we vary each of
these parameters in turn and show the Mach number and $v_z$ in
Fig.~\ref{fig:parameters}. We choose the simulation time such
that the jet has travelled approximately 150kpc in each case. 
The parameters chosen can change the aspect ratio and advance speeds,
but in each case fast supersonic backflows form and we observe a 
similar qualitative behaviour to the F1 simulation. The length
and width of the backflow is also affected by varying these 
parameters, but the prevalence of backflows is not limited to
a specific case. In the next section, we will discuss the 
physical conditions under which we expect backflows to form, 
with further discussion of the advance speed and jet morphology. 

\section{Dynamics and morphology of the jet, backflow and lobe}
\label{sec:backflows}
Backflows in jet cocoons have been discussed in analytic and
self-similar models \citep{falle_self-similar_1991,scheuer_lobe_1995}
and observed in the earliest jet simulations 
\citep{norman_structure_1982}. The strength of the 
backflows in the simulations of \cite{norman_structure_1982} may 
be artificially enhanced by adopting 
outflow boundary conditions at $z=0$ outside the 
jet nozzle \citep{koessl_1988}, but backflows nonetheless persist in a number of 3D 
HD and MHD simulations with more realistic boundary 
conditions \citep{saxton_complex_2002,gaibler_very_2009,
mathews_dynamics_2012,mathews_first_2014,cielo_3d_2014,tchekhovskoy_three-dimensional_2016}.
The presence of strong backflows in observations has
also been inferred in a number of studies
\citep{laing_relativistic_2012,mathews_first_2014}. Here, we examine
the physical requirements for backflow and compare them to constraints
from the jet morphology and dynamics. 

\subsection{When do backflows form?}
\label{sec:bernoulli}
We can gain some insight into the behaviour of the plasma in
the backflowing region by considering 
Bernoulli's principle, as applied to backflows from jets by 
\cite{norman_structure_1982} and \cite{williams_numerical_1991}.
Following \cite{williams_numerical_1991}, we consider
a fluid parcel that has passed through the termination shock and 
has a velocity $v_h$, density, $\rho_h$, and a pressure, $P_h$, directly 
downstream of the shock, set by the shock jump conditions. 
The fluid parcel is funnelled sideways and backwards away from the
hotspot. We now assume steady flow and consider the velocities
in the frame of the termination shock (which is moving slowly in the
observer frame if the density contrast is high). Neglecting gravity, we 
can use the non-relativistic steady-state momentum equation, 
$\bb{v} \cdot \nabla \bb{v} = - \nabla P / \rho$,
and make use 
of the identity $\bb{v} \cdot \nabla \bb{v} = \nabla (v^2/2) -
\bb{v} \times (\nabla \times \bb{v})$ and the adiabatic condition $P \rho^{-\gamma}=P_h \rho_h^{-\gamma}$
to write 
\begin{equation}
\nabla \left( \frac{v^2}{2} 
+ \frac{\gamma}{\gamma - 1} \frac{P_h^{1/\gamma} P^{ (\gamma-1)/\gamma}}{\rho_h} \right)
- \bb{v} \times (\nabla \times \bb{v}) = 0.
\end{equation}
We can now integrate along a streamline of steady flow 
to give a conserved, Bernoulli-like quantity
\begin{equation}
\chi = v^2 + 
\frac{2\gamma}{\gamma - 1} \frac{P_h^{1/\gamma} P^{ (\gamma-1)/\gamma}}{\rho_h}
= v_h^2 + 
\frac{2\gamma}{\gamma - 1}\frac{P_h}{\rho_h}.
\label{eq:backflow_velocity1}
\end{equation}
Under these assumptions the velocity $v$ will be maximum at a point 
along the streamline where $P$ is minimum.
Thus, the backflow is maximised when the pressure difference between 
the cocoon and hotspot is highest. The jet is confined by the pressure
in the cocoon and so is in rough pressure equilibrium with the cocoon 
far from the hotspot. We can therefore write an equation for the 
characteristic velocity of the backflow, $v_b$, 
under the additional assumption that $P\approx P_j$,
\begin{equation}
v_{b} = \left[ 
v_h^2 + \frac{2\gamma}{\gamma - 1} \frac{P_h}{\rho_h} - 
\frac{2\gamma}{\gamma - 1}
\frac{P_h^{1/\gamma} P_j^{ (\gamma-1)/\gamma}}{\rho_h}
\right]^{1/2}
\label{eq:backflow_velocity2}
\end{equation}
since $P_h$ is set by the termination shock jump conditions
and is proportional to $\rho_j v_j^2$, 
the backflow speed is maximised when the jet Mach number is high.
If we now for simplicity assume non-relativistic jump
conditions at the termination shock and set $\gamma=5/3$ 
we have $P_h = 3 \rho_h v_h^2$ and therefore 
the Mach number of the flow is given by
\begin{equation}
M^2 = \frac{1}{5} + \frac{16}{5} \left[ \left( \frac{P}{P_h} 
 + 1 \right) ^ {-2/5} \right].
\label{eq:backflow_mach}
\end{equation}
In this (illustrative) non-relativistic 
limit, the flow goes supersonic when $P \approx 0.57 P_h$ 
and $v \rightarrow v_j$ if $P \rightarrow 0$. 

The above analysis shows that as long as the pressure
in the stream is allowed to drop below a 
critical value then the flow must go supersonic. 
The pressure in the backflowing stream is governed 
by the pressure variations in the turbulent lobe, so there
is a complicated interaction between the collimation 
of this stream and the surrounding medium. 
The speed and Mach number of the backflow
are both maximised when the jet Mach number is high and 
when the jet is light with respect to its surroundings, 
as shown by other authors 
\citep[e.g.][]{norman_structure_1982,williams_numerical_1991}. 
Once the backflow is supersonic, the only way it can slow down 
is via shocks; hence, shocks are an inevitable feature of backflows in 
astrophysical jets. 

The backflows shown in the $x-y$ slices in Fig.~\ref{fig:slices} are supersonic, fast and radially thin compared to $r_j$. This behaviour is expected. Since much of the jet material is 
funneled along the backflow, conservation of mass in 
rough cylindrical symmetry gives
$\pi r_j^2 \rho_j v_j \sim  
2\pi r_b w_b \rho_b v_b$,
where $r_{b}$ is the radial distance of the 
backflow from the jet axis and $w_{b}$ is the 
radial width of the backflowing stream. 
Since $r_b > r_j$, as the pressure drops and $v_b$ increases the backflow becomes a thin, supersonic stream along which a large fraction of the jet's kinetic energy flux can be focused. 

\subsection{Advance speed and aspect ratio}
\label{sec:morphology}
Two important empirical measurements that place
constraints on jet physics are the advance speed of the  jet head,
$v_\mathrm{head}$, and the aspect ratio of the jet width to 
cocoon width, $A=R_c / R_j$. Advance speeds of FRII sources are generally
much lower than the jet velocity, often on the order of $0.01c$ for 
relativistic jets \citep[e.g]{scheuer_lobe_1995,carilli_cygnus_1996,blundell_nature_1999}.
The advance speed of the jet head, $v_\mathrm{head}$ 
is governed by the jet velocity $v_j$ and the relativistic jet density 
ratio, $\eta_r$, and from 1D ram pressure balance
one obtains \citep{marti_morphology_1997}
\begin{equation}
v_{\mathrm{head}} 
= \frac{\sqrt{\eta_r}}{\sqrt{\eta_r}+1} v_j.
 \label{eq:advance_speed}
\end{equation}
The relativistic generalisation of the jet density ratio increases rapidly for high 
$\Gamma_j$ ($\eta_r \propto \Gamma_j^2$, see equation~\ref{eta-rel}). 
Thus, it is quite difficult to 
arrange that a steady, light jet has high kinetic power but a 
relatively slow advance speed. In reality, the 
intermittency of the jet may be crucial in delivering
power down the jet nozzle without the average advance speed 
increasing dramatically, but a study of this is beyond 
the scope of this paper. 

We can also estimate the physical dependence of $A$, which will allow us to check if backflows should only occur in jets with certain morphologies. 
In FRII radio galaxies this ratio is generally quite large \citep{williams_numerical_1991,krause_very_2005,english_numerical_2016}
-- a ratio of $\sim30$ can be inferred from radio images of Cygnus A \citep{perley_jet_1984}.  
An approximate estimate of $A$ can be derived for a uniform ambient medium
if we model the cocoon as a cylinder of radius $R_c$ expanding in length 
at a rate $v_{\mathrm{head}}$, then the rate of energy change in this cylinder
is $\pi R_c^2 v_{\mathrm{head}} U_c$, where we assume a constant internal energy
density $U_c$. If we equate this to the jet power $Q_j$ 
from equation~\ref{eq:jet_power} and make the additional assumption that 
$P_j = P_c$ then we obtain 
\begin{equation}
A = \frac{R_c}{R_j} = 
\left[ 
\frac{
\Gamma_j (\Gamma_j-1) \rho_j c^2 
+ \frac{\gamma}{\gamma - 1} \Gamma_j^2 P_j
}
{2 (\gamma - 1) P_j}
\frac{v_j}
{v_\mathrm{head}}
 \right]^{1/2},
 \label{eq:aspect_ratio}
\end{equation}
which makes it clear that wider cocoons compared to the jet width are 
preferentially produced by slow advance speeds and high Mach number jets, as shown by
\cite{williams_numerical_1991}. Even if the assumption of $P_j = P_c$ is dropped,
large values of $A$ still occur when the pressure in the jet hotspot is large
compared to the average pressure in the cocoon. Intermittency, or ``dentist-drill'' 
variability \citep{scheuer_morphology_1982}, can also increase this aspect ratio, 
as has been shown in simulations of intermittent jets
\citep[e.g.][]{tchekhovskoy_three-dimensional_2016}.

Overall, the empirical evidence points towards wide cocoons
and slow advance speeds, which favours light, high Mach number jets 
(equations \ref{eq:advance_speed} and \ref{eq:aspect_ratio}).
These are also the conditions in which strong 
backflows are produced (equation~\ref{eq:backflow_velocity2}), 
which, together with the ubiquity of backflows for all 
parameters explored in section~\ref{sec:param_sensitivity},
allows us to conclude that backflows are not unique 
to our particular parameter space but instead should 
exist in a large fraction of powerful extragalactic radio sources.

\subsection{The effect of magnetic fields}
Neither our HD simulations, nor the Bernoulli argument above, accounts for the effects of the magnetic field. The magnetic field helps determine the maximum CR energy (see section~\ref{sec:magnetic}) but can also affect the jet confinement and the dynamics of the lobe. MHD simulations of AGN jets show various behaviours depending on the field topology and magnetization parameter $\sigma$, which is the ratio of the Poynting flux to kinetic energy flux in the jet frame. High $\sigma$ simulations can suppress backflow, instead forming a `nose cone' of material being collected ahead of the termination shock \citep{clarke1986,komissarov1999,gaibler_very_2009}, although this does not occur for purely poloidal fields \citep{leismann2005}. However, while the jet is expected to be Poynting flux dominated near the jet base \citep[e.g.][]{beskin2011,zdziarski2015}, such jets will likely become kinetically dominated beyond kpc scales \citep{appl1988,sikora2005}. A transition to kinetic energy dominance can be caused by the magnetic kink instability \citep{appl2000}, as shown by \cite[e.g.][]{giannios2006,tchekhovskoy_three-dimensional_2016}. Furthermore, observations of radio galaxies such as Cygnus A and Pictor A show extensive cocoons rather than a nose cone morphology \citep[e.g.][]{perley_jet_1984,hardcastle2005}, suggesting backflows are present. 

In lower $\sigma$ simulations, which produce a good match to observations \citep{hardcastle_numerical_2014}, the magnetic field can still affect the jet and lobe dynamics. \cite{gaibler_very_2009} find that helical magnetic fields in the jet can alter the advance speed (which could lead to lower values of $A$ than from equation~\ref{eq:aspect_ratio}), damp Kelvin-Helmholtz instabilities and widen the jet head. \cite{keppens2008} find interesting behaviour in the backflow region, where a compressed magnetic field between the backflow and jet can suppress the interaction between the two. Magnetic confinement of the jet \citep{begelman1984} may also alter the Bernoulli argument above, and change the partitioning of energy densities in the lobe needed to confine the jet by a numerical factor depending on the field topology. Overall, the dynamic impact of the magnetic field on the backflow certainly warrants further attention (see also section~\ref{sec:magnetic}), but is unlikely to change our general arguments.

\section{Shock properties}
\label{sec:shock_properties}

To examine shock properties we use an Eulerian method to calculate the 
shock size and Lagrangian tracer particles to measure the distribution of
shock Mach numbers and velocities that fluid elements pass 
through. We limit the range of times within which we analyse 
shocks so that the first time stamp analysed corresponds to
when the jet has advanced approximately $50$~kpc. This 
results in time ranges of $35.9$-$122.4$~Myr,
$6.2$-$40.8$~Myr and $6.2$-$32.64$~Myr for the S1, F1
and F3D simulations, respectively. All histograms
relating to Lagrangian tracer particles are normalised to the number 
of tracer particles injected during these time ranges, which we denote 
$N_{\mathrm{p,tot}}$.

\subsection{Eulerian shock statistics}
\label{sec:eulerian_shocks}
To calculate shock size $r_{\rm s}$, we first flag grid cells as inside shocks if they satisfy the conditions $\tilde{\nabla} P/P>0.2$ and $\nabla \cdot \boldsymbol{v}<-0.05 (c/\mathrm{kpc})$. We also impose the additional constraints of $C_j>0$ and $x>2$, in order to focus on the jet lobe and cocoon. We then make use the \verb|scikit-learn| \citep{scikit-learn} implementation of the Density-Based Spatial Clustering of Applications with Noise (DBSCAN) algorithm \citep{ester_density-based_1996} to identify shock regions. We call the \verb|cluster.DBSCAN| function in \verb|scikit-learn|, setting \verb|min_samples=5| and \verb|eps=2|. This means that the smallest number of grid cells that constitute a cluster is 5, while the shortest distance between two points in a cluster is twice the grid scale. 

\begin{figure}
\centering
\includegraphics[width=\linewidth]{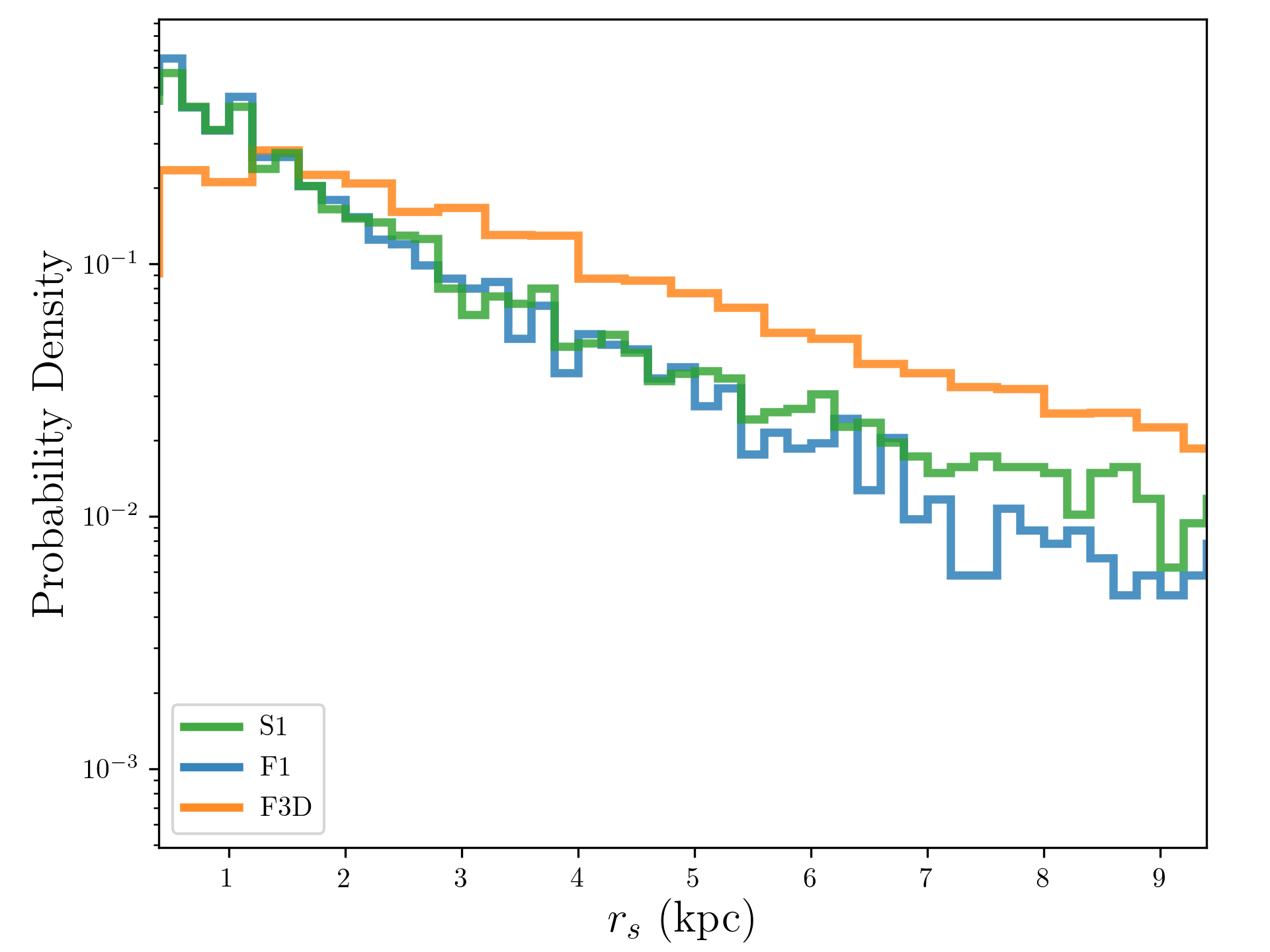}
\caption{Histogram of shock sizes in kpc 
from the S1, F1 and F3D simulations, as described in 
section~\ref{sec:eulerian_shocks}.
The histogram is in units of probability density.
}
\label{fig:sizes}
\end{figure}

Once the shock clusters have been identified, we  calculate the shock size by measuring the linear extent of each identified cluster and assuming (conservatively) a straight shock. A histogram of shock sizes is shown in Fig.~\ref{fig:sizes}. We find shock sizes ranging from just above the grid resolution up to nearly $10$kpc. For the 2D simulations, we only calculate the shock size in the $x-z$ plane rather than measuring the cylindrical extent of the shock, although in 3D the shock size is calculated in 3D and the slightly larger shock size is indicative of the partial ring shapes formed by the backflowing streams shown in Fig.~\ref{fig:slices}. The mean shock sizes in the S1, F1, and F3D simulations are $2.01$~kpc, $1.85$~kpc and $4.61$~kpc, respectively; we therefore take $2$~kpc as a typical shock size. 

\subsection{Tracer particle histories}
\label{sec:tracers}

\begin{figure}
\centering
\includegraphics[width=1.0\linewidth]{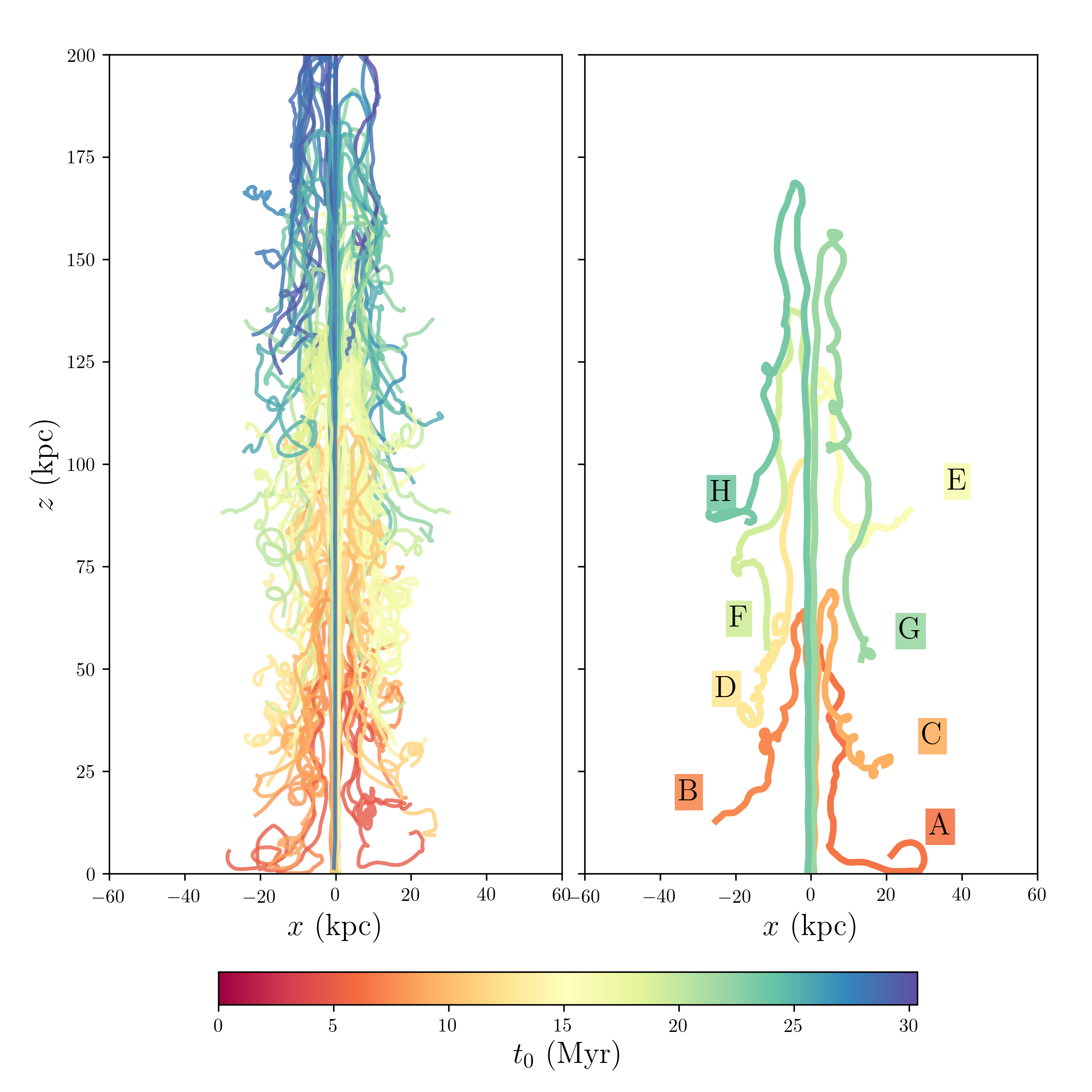}
\caption{{\sl Left:} Tracer particle trajectories from the F3d simulation
for 100 random tracer particles, with the sign of the $x$-coordinate 
also chosen randomly. The colour corresponds to the launch time of the 
particle, $t_0$, in Myr. {\sl Right:} Trajectories for 8 individual
tracer particles whose histories are shown in Fig.~\ref{fig:tracers}. The 
colour scheme is identical to the left panel and the labeled letters 
match those in Fig.~\ref{fig:tracers}.}
\label{fig:trajectories}
\end{figure}

Tracer particles are injected in the 2D simulations
at the jet nozzle as described in section~\ref{sec:shock_detect}. 
We record the local fluid properties for each particle as it is 
advected with the flow, writing to file every $\Delta t = 3,264$~yr (1 simulation time unit). 
The trajectories of 100 random tracer particles are shown in the left-hand panel of Fig.~\ref{fig:trajectories}, colour-coded by launch time, showing that the tracer particles propagate along the jet before invariably being channeled along backflowing streams. The vorticity in the backflow and further into the cocoon manifests as loops and twisting patterns in the trajectories of the particles. 

\begin{figure*}
\centering
\includegraphics[width=1.0\linewidth]{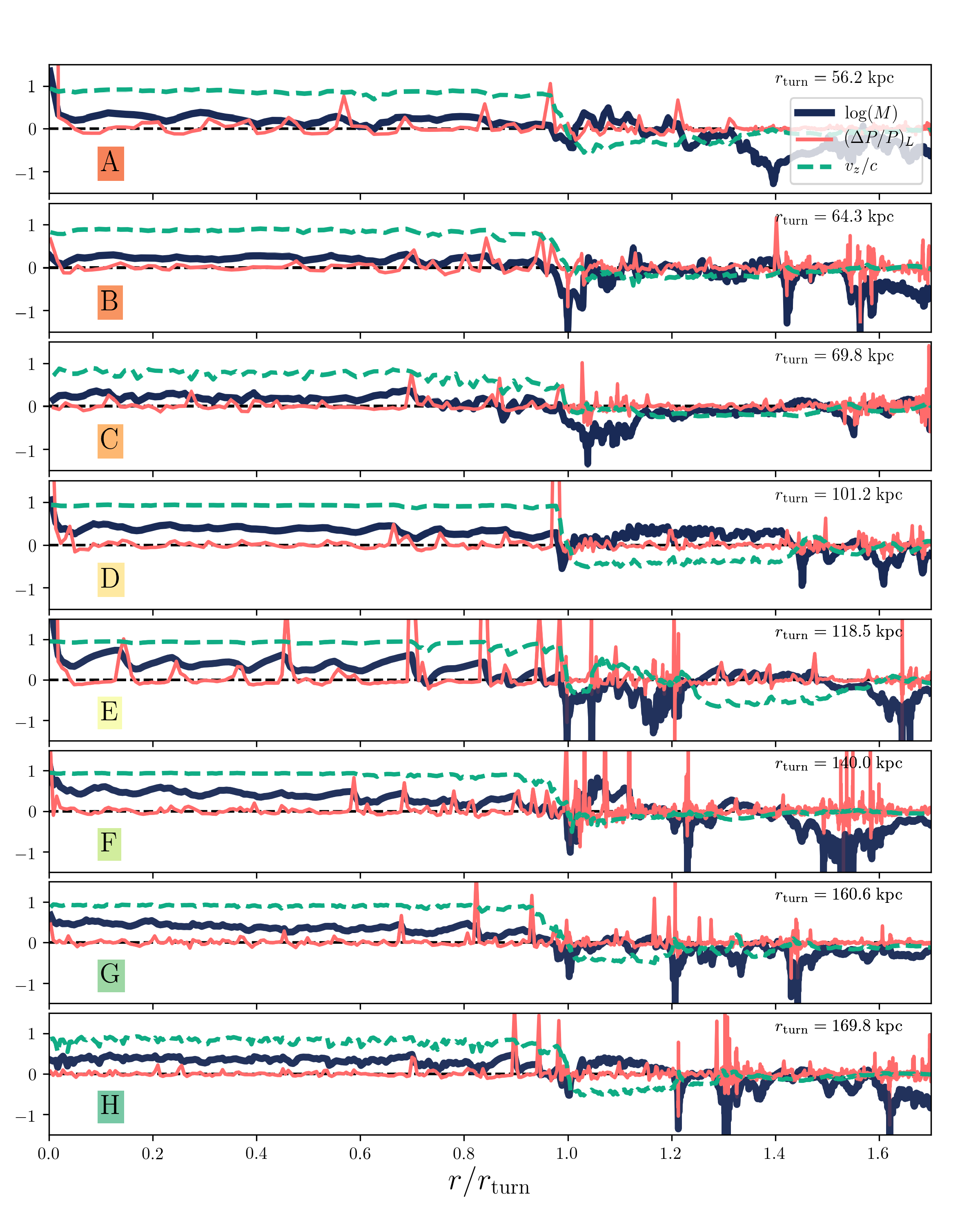}
\caption{Profiles of $\log(M)$, $(\Delta P / P)_L$  and $v_z$ 
for eight Lagrangian tracer particles in the F1 (2D) simulation. The distances
travelled by the particles, $r$, are normalised to the distance travelled at the 
first point when the $z$-component of the velocity of the local fluid has become negative 
($r_{\mathrm{turn}}$) -- the ``turning point'' 
after which the particle is then travelling along the backflow. 
While $v_z$ illustrates when the particle enters the backflow, 
$\log(M)$ and $(\Delta P / P)_L$ show supersonic flow and shocks. 
The labels in each panel match those in Fig.~\ref{fig:trajectories}.}
\label{fig:tracers}
\end{figure*}

The right hand panel of Fig.~\ref{fig:trajectories} shows trajectories 
for 8 individual tracer particles, while profiles of some key quantities as a function of distance travelled by the same tracer particles are shown in Fig.~\ref{fig:tracers}. We show the logarithm of $M$ and linear profiles of 
$(\Delta P / P)_L$ and $v_z$, with distance travelled normalised to $r_{\mathrm{turn}}$, the distance at which the sign of $v_z$ first becomes negative. The particles travel along the jet and pass through a series of reconfinement shocks. These reconfinement shocks can be seen in the colormaps in section~\ref{sec:results} and they show up as clear spikes in $(\Delta P / P)_L$  in the tracer profiles. This acts as a verification of this quantity as a shock diagnostic.  The reconfinement shocks cause the internal Mach number to drop, as expected, although the fluid generally remains supersonic, with $v_z$ close to its initial value of $0.95c$, until the jet terminates. At this point, there is usually a clear drop in $v_z$ and $M$. Shortly after this point, the sign of $v_z$ can become negative showing that the tracer particle has entered a backflowing stream. The Mach number can increase again in the backflow, and subsequent shocks are often encountered. After a time, the material is advected deep into the cocoon and comes into rough pressure equilibrium with the larger-scale surroundings. Turbulence and vorticity persist throughout this cocoon, but far from the hotspot the flow is generally subsonic, although occasionally transonic (see also Fig.~\ref{fig:m_and_dv}).

\begin{figure*}
\centering
\begin{subfigure}{.5\textwidth}
  \centering
  \includegraphics[width=\linewidth]{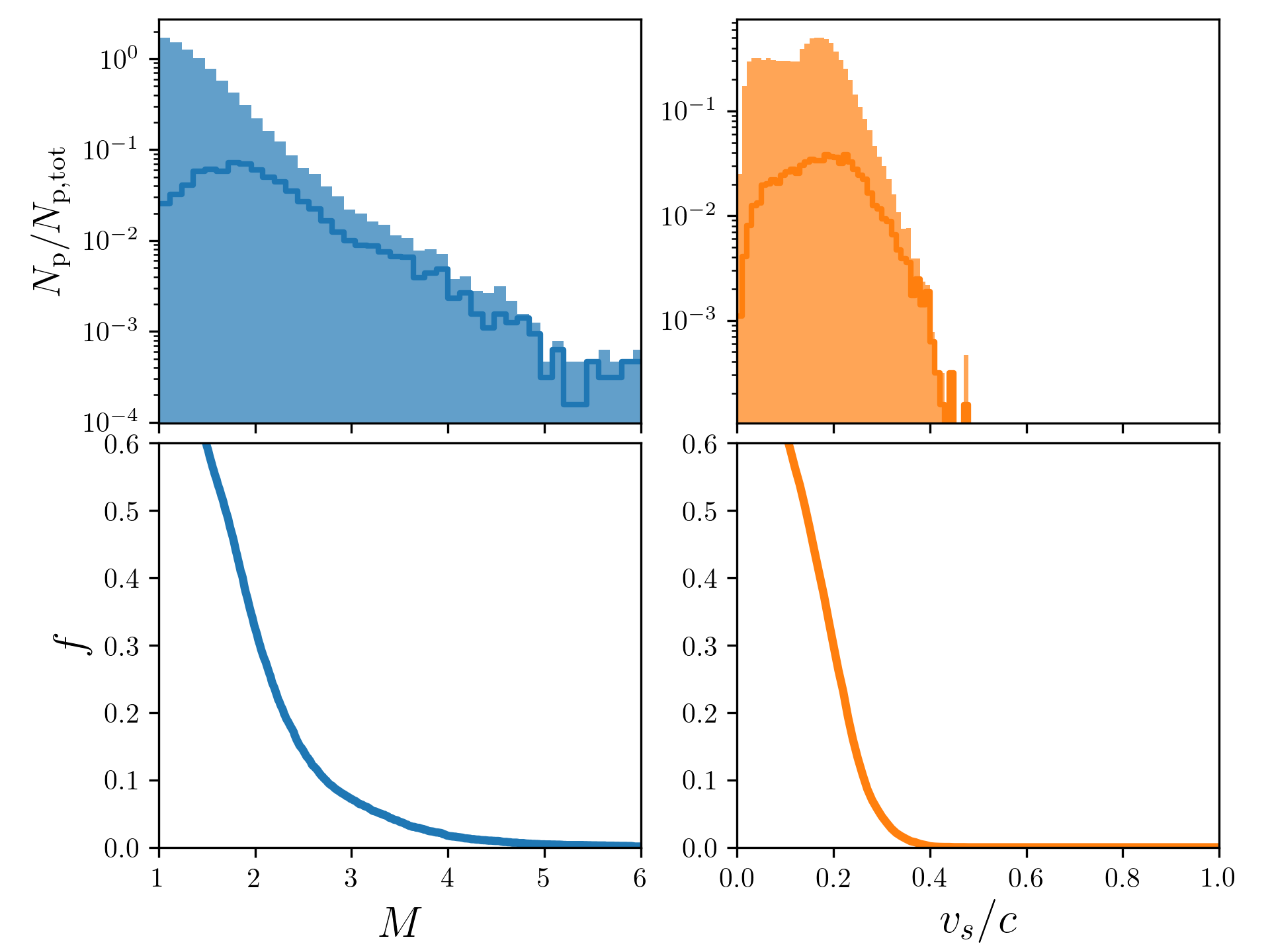}
  \caption{Shock properties in run S1}
  \label{fig:sub1}
\end{subfigure}%
\begin{subfigure}{.5\textwidth}
  \centering
  \includegraphics[width=\linewidth]{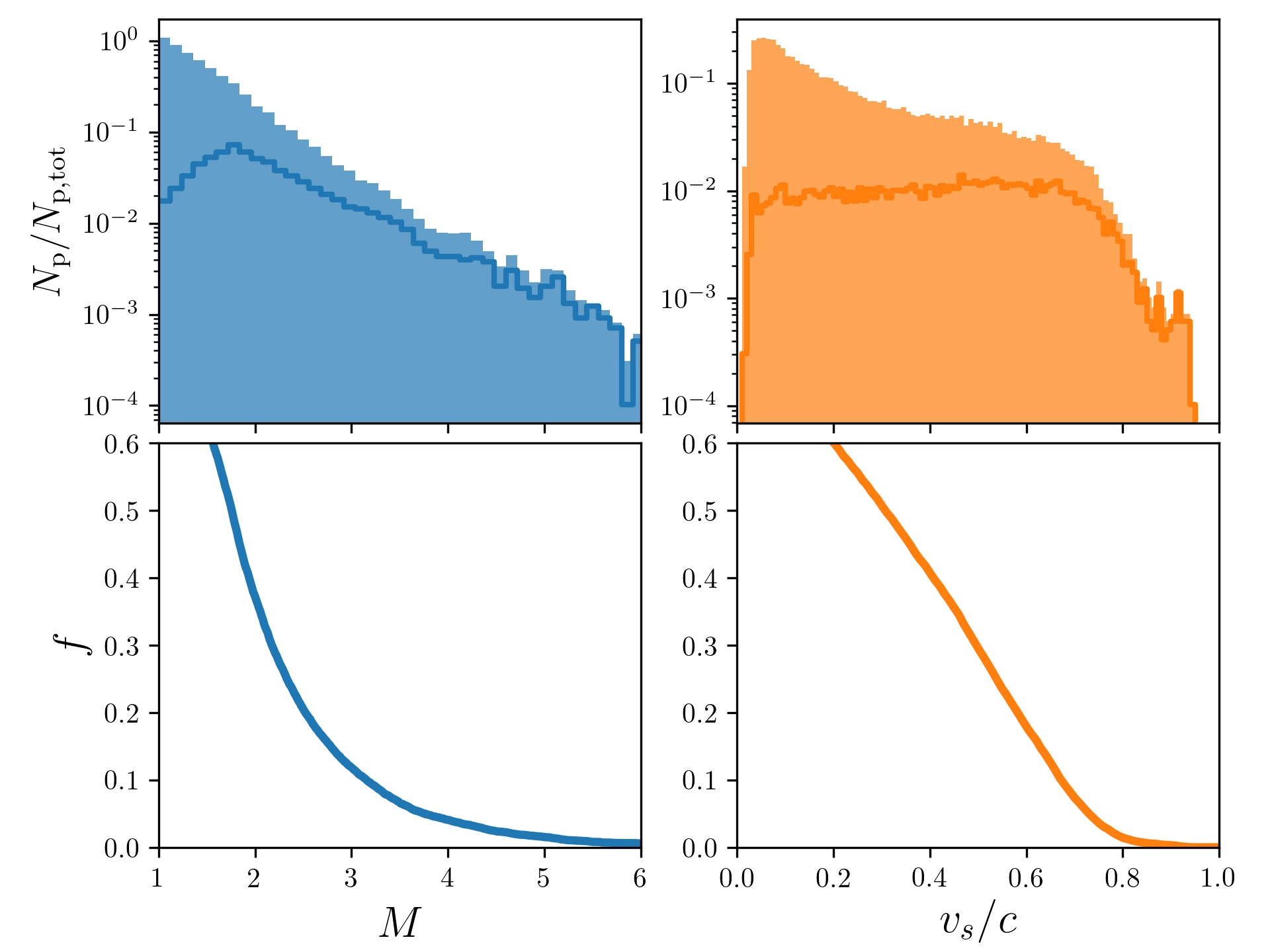}
  \caption{Shock properties in run F1}
  \label{fig:sub2}
\end{subfigure}
\caption{Shock Mach numbers and velocities as recorded by the Lagrangian tracer particles for both of the 2D simulations (S1 and F1). Top two panels: histograms of $M$ (left) and $v_s/c$ (right) 
the shocks passed through by all tracer particles within the time range considered. The solid histogram shows all shocks, while the solid line shows 
only the strongest shock that each particle passes through.
Bottom left: The fraction of particles passing through a shock with Mach 
number as least as high as $M$; this is equal to one minus the cumulative 
distribution function (CDF) of the solid line in the panel above. 
Bottom right: The fraction of particles 
whose strongest shock has a velocity of at least $v_s$.}
\label{fig:fractions}
\end{figure*}

\subsection{Lagrangian shock statistics}
\label{sec:lagrangian}

The tracer particle histories shown in Figs.~\ref{fig:trajectories} and 
\ref{fig:tracers} give some idea
of the shocks a fluid element might pass through in the backflow, but it is important to analyse this data statistically. Specifically, we are concerned with the percentage of tracer particles that pass through strong shocks with the right kind of shock velocities. We do this by identifying shocks as described previously by requiring $(\Delta P / P)_L<0.2$, $\nabla \cdot \boldsymbol{v}<-0.05 (c/\mathrm{kpc})$, $C_j>0$ and $x>2$. We also only record the shocks that have occurred once the tracer particle has entered the backflow, i.e. after the $z$-component of the velocity of the local fluid has first become negative (see Fig.~\ref{fig:tracers}). We record the properties ($M, v_s$) of each shock that each tracer particle passes through. The most important shock is the strongest one, since that will tend to have the flattest spectrum \citep{blandford_particle_1987} and will therefore dominate the UHECR contribution. 

In Fig.~\ref{fig:fractions} we show statistics for the Lagrangian tracer
particles in both our 2D runs. We give the distributions of $M$ and $v_s$ 
for all shocks that the tracer particles pass through, as well as just the
strongest shocks (highest $M$). We also show the fraction of particles
that have passed through a shock at least as strong as $M$, which is 
equal to one minus the cumulative distribution function of the histograms
in the top panel. These figures illustrate that in both cases approximately 
$10\%$ of particles pass through a shock of $M>3$. These shocks have a range 
of shock velocities and can be non-relativistic or mildly relativistic; 
we take $v_s=0.2c$ as a typical shock velocity. 

\subsubsection{Number of Shock Crossings by a fluid element}

\begin{figure}
\centering
\includegraphics[width=\linewidth]{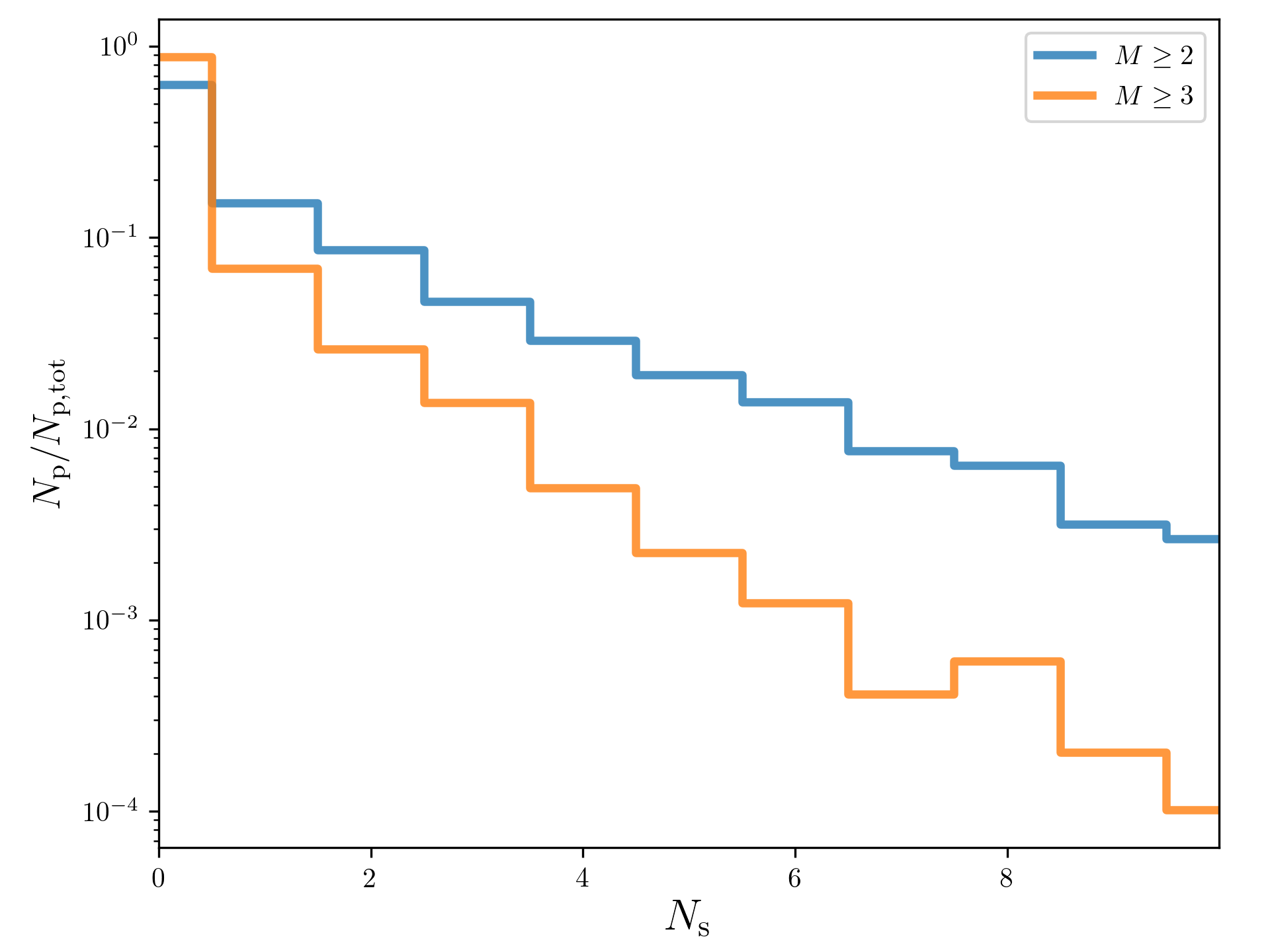}
\caption{A histogram showing the number of shock crossings, 
$N_{\mathrm{s}}$ by a tracer particle in run F1, for two 
different shock Mach numbers. $N_\mathrm{p}$ is the number of particles passing 
through $N_{\mathrm{s}}$ shocks, while $N_{\mathrm{p,tot}}$ is the total 
number of tracer particles injected in the simulation.}
\label{fig:crossings}
\end{figure}

Multiple shocks occur along the backflow and throughout the cocoon, as can be seen in e.g. Figs~\ref{fig:m_and_dv}. There is thus opportunity for fluid elements to cross multiple shocks. Fig.~\ref{fig:crossings} shows a histogram of the number of shock crossings by the Lagrangian tracer particles in the F1 simulation, for two different Mach numbers. The histogram is normalised so that it shows the fraction of all tracer particles passing through $N_{\mathrm{s}}$ shocks. The tracer particles often pass through more than one additional shock downstream of the jet termination shock, as would be expected from the backflows seen in Figs~\ref{fig:m_and_dv} and \ref{fig:3d_mach}. The percentage of particles passing through two or more $M>3$ shocks in the F1 simulation is $4.96\%$, compared to the $11.8\%$ of particles that pass through at least one $M>3$ shock.

A particle passing through a number of shocks can be further accelerated and the final CR-spectrum is harder than in a single shock acceleration \citep{bell_acceleration_1978-1,blandford_supernova_1980,achterberg_particle_1990,pope_diffusive_1994,melrose_effect_1997,marcowith_computation_1999,gieseler_first_2000}. The situation in the backflow is therefore similar to that considered by \cite{meli2013}, except that their analysis concerns oblique reconfinement shocks in the jet. Multiple shock crossings make the overall conditions favourable for acceleration to high energy, as not only can existing CRs be further accelerated but the magnetic field has multiple opportunities for amplification. We discuss the latter further in section~\ref{sec:magnetic}. Concerning the maximum energy of particles, the upper-limit is still set by the size of the shocks and the value of the magnetic field, as we describe in the next section, but $N_s$ shock crossings will make conditions more favourable and increase the maximum CR energy by a factor on the order $N_s$.

\section{Maximum Cosmic-Ray Energy}
The Hillas energy, given by Eq.~(\ref{E-Hillas}),
sets the characteristic maximum energy 
achievable by a CR. To estimate $E_H$ we adopt values of
$v_{\rm s} = 0.2c$ and $r_{\rm s} = 2$~kpc informed by the 
results of the previous 
section, but the appropriate value of the magnetic field 
$B$, as well as the composition of UHECRs (and therefore 
appropriate $Z$) is also crucial. 

\subsection{Magnetic Field}
\label{sec:magnetic}
Our simulations are HD, rather than MHD, so we do not solve the 
induction equation. The reasoning for this is partly that the magnetic 
field that matters for accelerating CRs to high energy is the 
turbulent, amplified field at the shock, which is small-scale until it 
grows to the Larmor radii of the highest energy 
CRs. Furthermore, the amplification is driven by streaming or drifting CRs 
with a spectrum of energies, which grow the turbulence  on a variety of 
scales. Instead of trying to resolve and self-consistently model the 
instabilities that amplify the field, we instead make some 
general arguments informed by plasma physics modelling and the 
acceleration of Galactic CRs by SNRs.

Turbulent magnetic field amplification is a general feature of any DSA theory \citep{bell_acceleration_1978,bell_acceleration_1978-1}.
Current-driven instabilities can amplify the field via a $\boldsymbol{j}_{\mathrm{ret}} \times \boldsymbol{B}$-force that stretches and distorts the field, where $\boldsymbol{j}_{\mathrm{ret}}$ is a return current produced in reaction to the CR current. At wavenumbers resonant with the Larmor radius of the CRs producing the current, this instability is known as the resonant or \alfven\ instability
\citep{lerche_unstable_1967,kulsrud_effect_1969,wentzel_cosmic-ray_1974,skilling_cosmic_1975,skilling_cosmic_1975-1,skilling_cosmic_1975-2}.
The resonant instability can only amplify the field to $\delta B / B \sim 1$ \citep[e.g.][]{amato_kinetic_2009,bell_particle_2014}. The non-resonant hybrid (NRH)  or Bell instability can amplify the magnetic field to many times its ambient value \citep{bell_turbulent_2004,zirakashvili_modeling_2008,niemiec_production_2008,stroman_kinetic_2009,riquelme_nonlinear_2009,bell_cosmic-ray_2013} 
in the case of both parallel and perpendicular initial field orientations \citep{bell_interaction_2005,milosavljevic_cosmic-ray_2006,riquelme_magnetic_2010,matthews_amplification_2017}.

\cite{bell_cosmic-ray_2018} showed that on small scales,
within one UHECR Larmor radius of a highly relativistic (quasi-perpendicular) shock, 
the magnetic field is not amplified on a scale length large enough to
scatter UHECRs. Similar arguments were applied to the hotspots of FRII radio galaxies, 
where the maximum energy of the electrons at the termination shock (hotspot)
is set by the growth of turbulence on the scale of a Larmor radius 
in the perpendicular unperturbed magnetic field
and not synchrotron cooling as was usually assumed \citep{araudo_evidence_2016,araudo_maximum_2018}.
The same constraint applies to protons and therefore the CR maximum energy 
is about 1~TeV,  the same as
the maximum electron energy inferred by the optical-IR synchrotron cutoff.
In the particular case of the FRII radio galaxy Cygnus~A, the magnetic 
field in the hotspot can be amplified to large values of 
$50-400~\mu$G, but not on the right scale for acceleration to EeV energies
\citep{araudo_maximum_2018}.

Secondary shocks in the backflows of the jets have a number of advantages 
over the termination shock in terms of their prospects for UHECR production.
These include:
\begin{enumerate}
\item they span a range of velocities and so they include non- and mildly relativistic shocks;
\item they occur after the termination shock and so can make use of the already 
amplified field as a small-scale seed field;
\item fluid elements can pass through multiple shocks providing multiple 
opportunities for acceleration and field amplification by 
CR streaming instabilities;
\item the magnetic field can be amplified by other mechanisms (e.g. vorticity)
and not only CR-driven instabilities over Larmor radii scale lengths 
at the shock.
\end{enumerate}
Point (i) is important in order to avoid the issues inherent to relativistic shocks described in detail by \cite{bell_cosmic-ray_2018}. The final three points are crucial in terms of allowing the magnetic field to grow to the right scale and strength to accelerate UHECRs. In SNRs, the maximum CR energy is generally much lower than the Hillas energy, since one must take into account effects from the CR diffusion coefficient and system age \citep{lagage_cosmic-ray_1983} as well as the timescales and saturation fields associated with the amplification mechanism at the shock \citep{bell_cosmic-ray_2013}. If the magnetic field is amplified by the NRH instability then it saturates once the $\boldsymbol{j} \times \boldsymbol{B}$ force from the CR return current is balanced by magnetic tension, that is when $\mu_0 \boldsymbol{j} \sim \nabla \times \boldsymbol{B}$. While this effect is not always restrictive \citep{bell_cosmic-ray_2013}, the effect of the magnetic energy density being spread over many decades in the scale size of the field can be. The result is that the fields measured in synchrotron observations of SNRs, typically $100$s of $\mu$G \citep{cassam-chenai_blast_2007,uchiyama_extremely_2007}, cannot be naively substituted into the equation for the Hillas energy. 

In jet backflows, the physical situation is fundamentally different to that in SNRs. \cite{giacalone_magnetic_2007} have shown that density perturbations, which are bound to exist in such a dynamic and variable environment, can cause the field to become amplified a long way downstream of the shock. This, along with the general vorticity expected in jet hotspots and their associated backflows \citep[e.g.][]{norman_structure_1982,falle_self-similar_1991} can stretch field lines, amplifying the field and allowing the scale length to grow to the Larmor radius of a UHECR. We therefore expect the maximum energy of particles accelerated in these backflow shocks -- and any secondary shocks downstream of the termination shock -- to be much closer to the Hillas energy than in SNRs. This is because {\em there is more than one mechanism, and more than one opportunity, for the field to be amplified and stretched on the scale of an UHECR Larmor radius.}

The qualitative picture we have outlined is physically motivated but requires further investigation. Much of the physics is ``subgrid'' level; a full treatment would require resolution spanning decades and is not feasible, although shock-tube style simulations designed to imitate a streamline might prove useful in terms of estimating the impact of vorticity and dynamo action on the turbulent field, as in e.g. \cite{giacalone_magnetic_2007}. We note that \citet{de_young_magnetic_2001,de_young_magnetic_2002} argues that field amplification must occur along a streamline from jet to lobe, as otherwise jet magnetic fields would need to be restrictively high if they were just passively advected into the lobe. For the purposes of our maximum energy estimate, we assume that the mechanisms described are sufficient to ensure that a relatively large fraction of the total available energy is transferred to the magnetic field in the vicinity of shocks. We therefore estimate a characteristic field strength in each shock region in our simulations using the formula 
\begin{equation}
\bar{B} = \sqrt{2\mu_0\eta_B~(U + 1/2 \rho v^2)}
\label{eq:b_bar}
\end{equation}
where $\eta_B$ is an efficiency parameter that we set to $0.1$. We record this value for the strongest shock that each tracer particle passes through. This produces mean values for $\bar{B}$ of $15.18\mu$G and $26.95\mu$G for the S1 and F1 runs, respectively.  

\subsection{UHECR composition and charge}
The composition distribution of UHECRs is still debated. 
Measurement of the distribution of $X_{\mathrm{max}}$, the depth 
at which CR-induced air-showers reach their maximum energy 
deposit, is the main composition diagnostic for observatories such 
as Telescope Array (TA) and PAO. TA results have suggested protonic composition at the highest energies \citep{abbasi_study_2015}.
However, fitting the $X_{\mathrm{max}}$ distribution is 
model-dependent, and a recent comparison of the TA and PAO 
datasets which attempts to account for the differences in
the detector chain and analysis finds that the results from the 
two observatories are consistent within systematic uncertainties
\citep{the_telescope_array_collaboration_pierre_2018}. 
These results are then compatible with the overall PAO 
results, which generally point towards a mixed composition of 
protons, intermediate nuclei and Fe 
\citep{pierre_auger_collaboration_depth_2014,de_souza_measurements_2017}, 
with the heavier elements becoming progressively more
important at higher energy as would be expected. 
A heavier composition at higher energy is also observed 
in Galactic CRs beyond TeV energies 
\citep[e.g.][]{mueller_energy_1991}.
Therefore, although we 
provide estimates of the Hillas energy from our simulations for 
a few different values of $Z$, we take acceleration of protons to 
$10^{19}$eV to be our criterion for success given the latest 
composition results. In terms of rigidity, ${\cal R} = E / (Ze)$, 
this is equivalent to ${\cal R} = 10 EV$.

\subsection{Our estimate of the maximum CR energy}

\begin{figure}
\centering
\includegraphics[width=1.0\linewidth]{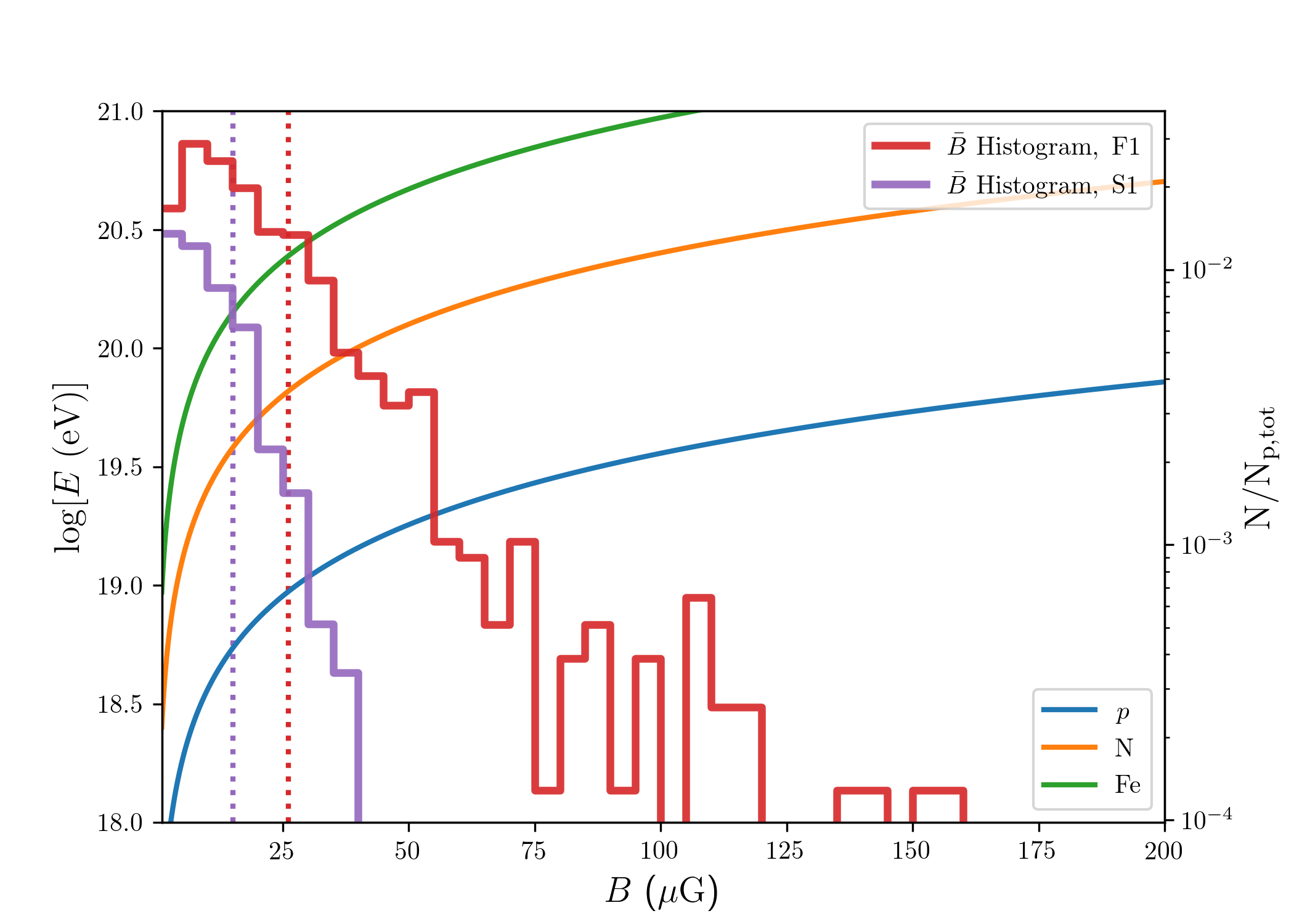}
\caption{Maximum (Hillas) energy  of protons, N and Fe 
for different values of $B$ (blue, orange and green curves)
and for our chosen representative shock parameters of $v_s=0.2c$ and $r_s=2$~kpc. 
The red histogram shows the distribution of $\bar{B}$ from equation~\ref{eq:b_bar}
for the strongest shocks that each tracer particle passes through in
the F1 and S1 simulations, while the dotted lines show the mean values.}
\label{fig:tmax}
\end{figure}

In Fig.~\ref{fig:tmax} we show curves of the Hillas energy for $r_s = 2$kpc, 
$v_s=0.2c$ for protons ($Z =1$, blue-solid line), He nuclei ($Z =2$, orange-solid 
line), and Fe nuclei ($Z =26$, green-solid line) as a function of the magnetic 
field. Overplotted is the distribution of $\bar{B}$ 
for the strongest shocks that each tracer particle passes through 
in the F1 and S1 simulations simulation, while the dotted lines show the mean values of these
histograms ($15.18\mu$G and $26.95\mu$G). For $B=26.95\mu$G, $v_s=0.2c$ and $r_s = 2$kpc, 
the Hillas energy is $9.70\times10^{18}$eV, while for the strongest magnetic 
fields ($\approx 140\mu$G) the Hillas energy is $5.04\times10^{19}$eV. 

\subsection{Scalability}
The quoted values for $\bar{B}$ and $E_{H}$ 
are for a specific simulation, but will scale with the physical parameters.
The magnetic field confining the CRs
should be proportional to $\sqrt{\rho_j v_j^2}$ \citep{bell_cosmic-ray_2013},
while the characteristic sizes in the system will scale with jet width, $r_j$,
provided that the value of $r_c$ is also scaled accordingly with $r_j$. 
If the most efficient acceleration to high energy always occurs at some critical 
shock velocity then we should expect $E_{\mathrm{max}} \propto \sqrt{\rho_j v_j^2} r_j$, that is,
the maximum energy should be proportional to the square root of the jet power,
although in reality the scaling is likely to be more complex. 
The jet powers we have adopted are in the FRII range but 
dramatically lower than estimates for the kinetic jet power in Cygnus A, for 
example \citep[$\sim 10^{46}$~erg~s$^{-1}$;][]{wilson_cavity_2006,ito_large_2007,kino_estimate_2005}, 
implying that maximum CR energies can be higher 
in certain sources. If non- or mildly relativistic shocks in backflows are 
indeed ubiquitous then the limiting factor on the CR energy is probably 
the jet power. Given this expectation, we discuss some general jet power 
requirements with reference to both the observed radio galaxy luminosity function 
and the kinetic power to radiative luminosity relationship in the next section. 

\subsection{Other types of shock}
In our analysis, we have focused on shocks in the lobes of radio galaxies, which are primarily produced in the supersonic backflows that form near the jet hotspot. It is interesting to consider whether the bow shock, reconfinement shocks, or termination shocks may prove to be good UHECR accelerators. The strongest shock in the jet-lobe system is typically the termination shock, but this is expected to be relativistic \citep[e.g.][]{begelman1984} and thus a poor accelerator to EeV energies \citep{lemoine_electromagnetic_2010,reville_maximum_2014,bell_cosmic-ray_2018}. It is possible that the termination shock velocity is not highly relativistic, in which case termination shocks may still be able to accelerate UHECRs, especially since the critical velocity below which acceleration to high energy becomes efficient is not yet clear. However, observations of radio galaxy hotspots suggest that jet termination shocks cannot accelerate particles to EeV energies \citep{araudo_evidence_2016,araudo_maximum_2018}. 
Jet reconfinement shocks have similar difficulties, since they are also generally relativistic and their oblique geometry lowers the energy gain per shock crossing \citep[e.g.][]{meli2013}. The complex interaction between the jet and the cocoon does produce some extended shock features, some of which will be included in the criteria for our shock detection as described in section~\ref{sec:shock_properties}. These structures merit future investigation but at face value appear less attractive UHECR accelerators than the shocks in backflows. 

The bow shock may also accelerate particles, but the shock velocity is low -- approximately equal to the jet advance speed ($\sim c/100$) at the tip of the bow shock and lower by a geometric factor away from the jet head. The bow shock smoothly transitions into a sound wave for the slowest advance speeds or high external pressures, as can be seen in Fig.~\ref{fig:parameters}. The magnetic field is also lower in the bow shock than in the jet or lobe as it is just a compressed version of the ambient field, while amplification is likely inefficient on UHECR Larmor radius scales. These factors may partly account for the absence of synchrotron emission from radio galaxy bow shocks \citep{carilli1988}, although synchrotron X-ray emission can be detected in Centaurus A's bow shock, where the maximum particle energy is thought to be well below the UHECR regime \citep{croston_high-energy_2009}. The bow shock in radio galaxies is likely a poor accelerator of UHECRs.

\section{Discussion}
\label{sec:discuss}

We have demonstrated that FRII-like jets produce the right kind of
shocks to accelerate CRs to rigidities above $10$EV, but this is 
not the only requirement for a successful model for UHECR production. 
We therefore discuss constraints on the power, isotropy, proximity 
and composition requirements for UHECR sources, informed by 
results from UHECR detectors and radio surveys and our recent
discussion of UHECR anisotropies \citep{matthews_fornax_2018_mnras}.
In this discussion, attenuation due to the 
GZK effect \citep[][]{greisen_end_1966,zatsepin_upper_1966}
and photodisintegration \citep{stecker_photodisintegration_1999} 
is important as it sets the characteristic maximum 
distance a high-energy proton or nucleus can travel. 
The horizon distance is strongly energy and composition dependent,
but is typically on the order of $100$Mpc 
\citep[e.g.][]{alves_batista_crpropa_2016,wykes_uhecr_2017}.
We adopt this as a canonical value but note the wide variation
in attenuation lengths for different species and at different CR
energies.

\subsection{Are there enough powerful sources?}
\label{sec:cr_power}

At least two basic energetic requirements must be satisfied by an UHECR source. The first is the {\em minimum power constraint}, described in various contexts by a number of authors
\citep{lovelace_dynamo_1976,waxman_cosmological_1995,waxman_high-energy_2001,blandford_acceleration_2000,massaglia2009}. For acceleration to a given rigidity, this constraint requires that sufficient power passes through the shock for the magnetic field to reach ${\cal R}/(v_s r_s)$. In the case of particle acceleration at shocks this can be computed by considering the magnetic energy density $U_{\mathrm{mag}} = B^2/(2 \mu_0)$ and the Hillas energy (equation~\ref{E-Hillas}). Since the maximum magnetic power delivered through a shock of size $r_s$ is approximately $U_{\mathrm{mag}} v_s r_s^2$, we can write an equation, independent of $B$, for the minimum power $Q_{\mathrm{min}}$, given by 
\begin{equation}
Q_{\mathrm{min}} = f_s~\frac{{\cal R}^2}{2 \mu_0 v_s}, 
\end{equation}
which is equivalent to
\begin{equation}
\left(\frac{Q_{\mathrm{min}}}{\rm erg\, s^{-1}}\right) \sim 
10^{44}
~\left( \frac{f_s}{0.1} \right)^{-1}
~\left( \frac{v_s}{0.1c} \right)^{-1}
~\left( \frac{{\cal R}}{10 \,\rm EeV} \right)^2.
\label{eq:power}
\end{equation}
Here, $f_s$ is the fraction of the jet's overall energy that is channeled through the right kind of shock for acceleration to high energy. We adopt $f_s = 0.1$ based on the results from section~\ref{sec:lagrangian}. 

The second energetic requirement for UHECR sources is that {\em the observed number of UHECRs arriving at Earth} can be produced. To calculate the luminosity in UHECRs that can be produced from a radio galaxy, we consider a jet of kinetic power $Q_j$, with some fraction $f_s$ of the jet power channelled through the right kind of shock for acceleration to high energy, and a further characteristic fraction, $\eta$ of each shock's energy budget going into UHECRs above energy $E_u$. In reality, $f_s$ and $\eta$ take different values for different shocks and for different values of $E_u$; nonetheless, we can estimate characteristic values. The value of $\eta$ can be estimated by considering a differential particle number distribution proportional to $E^{-\beta}$ in the number of CRs such that the differential luminosity is $dL/dE \propto E^{1-\beta}$. If we adopt $E_1=1$~GeV ($\sim m_p c^2$) as the lower energy bound for the total CR luminosity and define $\eta_0 = 0.3$ as the fraction of the shock energy going into CRs at all energies then we obtain 
\begin{equation}
    \eta(E>E_u) =
    \begin{cases}
      \eta_0 \frac{\ln (E_2/E_u)}{\ln (E_2/E_1)} \approx 0.05, 
      & \text{if}\ \beta=2, \\
      \eta_0 \frac{E_2^{-0.2} - E_u^{-0.2}}{E_2^{-0.2} - E_1^{-0.2}} \approx 0.003 & \text{if}\ \beta=2.2, \\
    \end{cases}
\end{equation}
where $E_2=10^{20}$~eV is a maximum energy cutoff and for the spectral index we choose as representative examples $\beta=2.2$ as expected for the intrinsic Galactic CR spectrum \citep{gaisser_gamma-ray_1998,hillas_topical_2005} and $\beta=2$ to match the theoretical expectation from DSA \citep{bell_acceleration_1978}. Clearly, the efficiency of UHECR production is strongly dependent on the CR spectral index and characteristics of the shock accelerating the CRs.

\begin{figure}
\centering
\includegraphics[width=\linewidth]{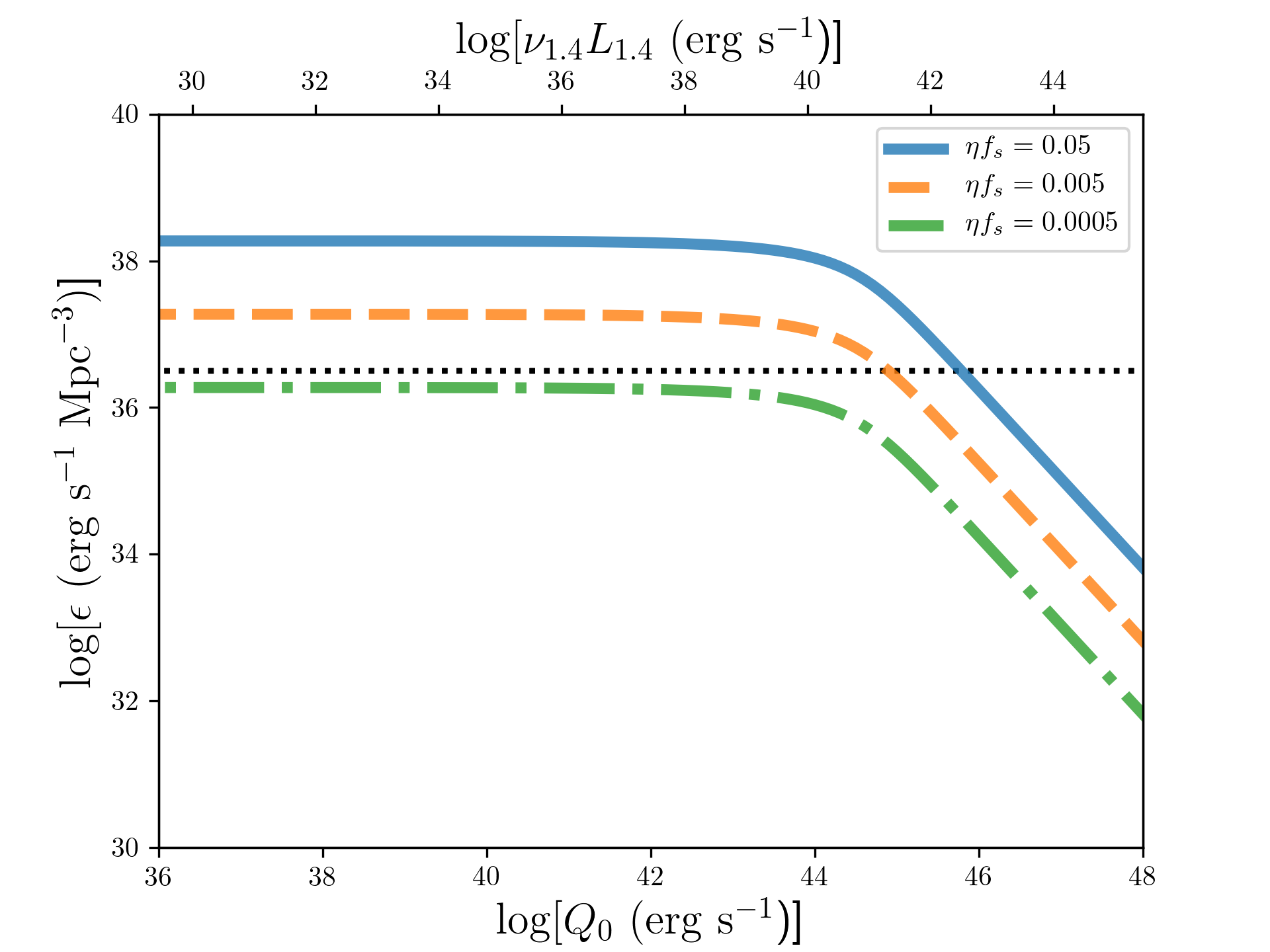}
\caption{The UHECR luminosity density produced by a population of radio galaxies as a function of the lower limit on the integral over jet power in equation~\ref{eq:uhecr_lum_density}, $Q_0$. The horizontal line shows the approximate observed UHECR luminosity density above $10^{18}$~eV observed at Earth. 
}
\label{fig:uhecr_lum_density}
\end{figure}

Jet power is related to the observed radio luminosity of a system; for the purposes of this estimate we adopt equation 1 from \cite{cavagnolo_relationship_2010}, which can be written as 
\begin{equation}
\left( \frac{Q_j (L_{1.4})}{\mathrm{erg~s^{-1}}} \right) = 
10^{0.75 \log (\nu_{1.4}~L_{1.4}) + 13.91}
\label{eq:cavagnolo}
\end{equation}
where $L_{1.4}$ is the monochromatic radio luminosity at $1.4$~GHz in erg~s$^{-1}$~Hz$^{-1}$. For a given 1.4 GHz luminosity function $\phi$ in units of Mpc$^{-3}\log(L_{1.4})^{-1}$, the luminosity density in UHECRs is 
\begin{equation}
\left( \frac{\epsilon_{u}}{\mathrm{erg~s}^{-1}~\mathrm{Mpc}^{-3}} \right)
= f_s \eta \int^\infty_{Q_{\mathrm{0}}} Q_j (L_{1.4})~\phi~d (\log L_{1.4})
\label{eq:uhecr_lum_density}
\end{equation}
We adopt the double power law luminosity function for radio galaxies given by \citet[][eq. 7]{heckman_coevolution_2014} with a break at $10^{31.95}$~erg~s$^{-1}$~Hz$^{-1}$ and plot $\epsilon_{u}$ as a function of $Q_{\mathrm{0}}$ in Fig.~\ref{fig:uhecr_lum_density}, for some representative values of $\eta~f_s$. Comparing to the dotted line, which shows the the UHECR luminosity density (luminosity per unit volume) above $10^{18}$eV reported by \cite{nizamov_constraints_2018} of $\epsilon_{18} = 10^{44}$erg~yr$^{-1}$~Mpc$^{-3}$, the figure makes it clear that, within the approximate spirit of the calculation, powerful radio galaxies are common enough to produce the observed UHECR fluxes at Earth. 
However, when we consider acceleration to $\sim 60$EeV and beyond the additional constraint of the GZK/photodistintegration horizon is important. As shown by a number of authors \citep[e.g.][]{blandford_acceleration_2000,massaglia_radio_2007,eichmann_ultra-high-energy_2018,matthews_fornax_2018_mnras}, powerful radio galaxies within $\sim100$~Mpc are scarce. 

\begin{figure*}
\centering
\includegraphics[width=\linewidth]{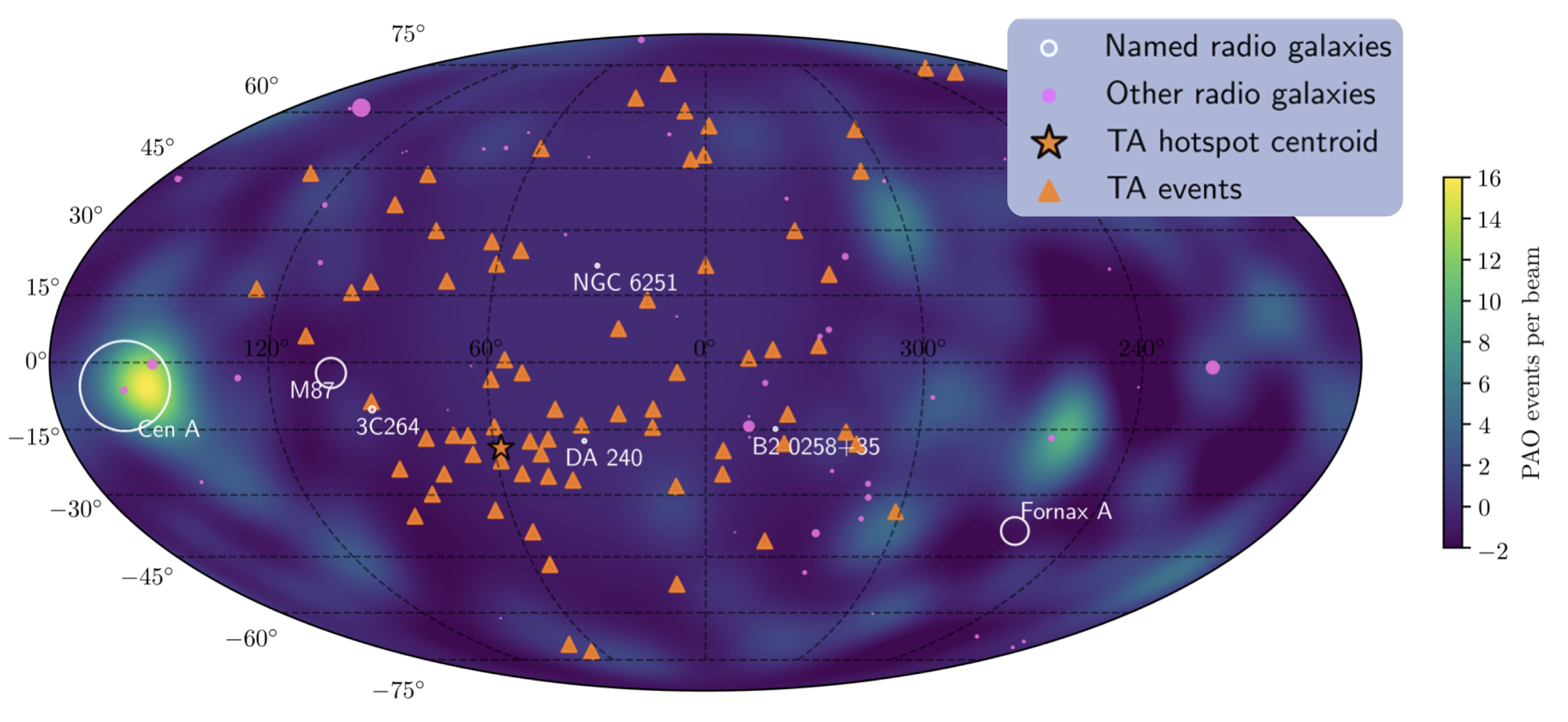}
\caption{A Mollweide projection in supergalactic coordinates showing a colormap of the PAO events per beam above $60$EeV, with the TA event ($>57$EeV) arrival directions overlaid (orange triangles) and the TA hotspot centroid from \protect\cite{abbasi_indications_2014} marked with a star. 
The circles show all radio galaxies from the vV12 catalogue with distances $<150$~Mpc and with radio 
luminosities $\nu L_\nu>2\times10^{40}$~erg~s$^{-1}$ at 1.4 GHz, with the size of the circles proportional to the observed flux density.
The white circles are specifically discussed in the text; otherwise the circles are colour pink. 
}
\label{fig:uhecr_map}
\end{figure*}

\subsection{UHECRs from dormant lobes inflated by powerful jets?}
\label{sec:directions}
In \cite{matthews_fornax_2018_mnras}, we showed that the excesses above isotropy in the PAO data \citep{pierre_auger_collaboration_indication_2018} may be explained by considering strong UHECR contributions from Centaurus A and Fornax A. Both these sources show giant lobes whose total energy contents are large compared to the energy input from the currently active jet in the system. There is also evidence of declining AGN activity in Fornax A \citep{iyomoto_declined_1998,lanz_constraining_2010} and of recent merger activity in both sources \citep{mackie_evolution_1998,horellou_atomic_2001}. Together, these considerations led us to invoke a scenario in which Fornax A and Centaurus A had both had more powerful, possibly FRII-like, jet ``outburst'' in the past, during which UHECRs were accelerated, meaning that their giant lobes now act as (slowly leaking) UHECR reservoirs. 

The observed UHECR excess map above $60$~EeV from  \cite{pierre_auger_collaboration_indication_2018} in supergalactic coordinates is shown as a Mollweide projection in Fig.~\ref{fig:uhecr_map}. The two ``hotspots'' discussed by 
\cite{matthews_fornax_2018_mnras} can be seen close to Fornax A and Centaurus A. The 
TA events above $57$~EeV from \cite{abbasi_indications_2014} are overlaid on the plot,
together with the positions of all radio galaxies from the \cite{van_velzen_radio_2012}
catalogue that are within 150~Mpc of Earth and have radio luminosities 
$\nu L_\nu>2\times10^{40}$~erg~s$^{-1}$ at 1.4 GHz. This luminosity cutoff corresponds to a minimum kinetic power of 
approximately $10^{44}$~erg~s$^{-1}$ 
(see equation~\ref{eq:cavagnolo}). 
The radio galaxies discussed in this paper are labelled.

The TA arrival directions have an excess just below the supergalactic plane, often referred to as the TA hotspot, with a characteristic spread of $\sim20^\circ$ \citep{abbasi_indications_2014}. It is possible that the TA events are dominated by a fairly diffuse component along the supergalactic plane, whereas the PAO events could instead be dominated by a few nearby radio galaxies in the southern sky. It might also be possible to explain the TA hotspot by an association with individual radio galaxies. The radio galaxy NGC 6251 is intriguingly similar to Centaurus A and Fornax A in that it has extremely large lobes (linear extent $2$ Mpc) with large total energy contents \citep{waggett_ngc_1977} and is thought to be an extended $\gamma$-ray source \citep{takeuchi_suzaku_2012}. 
Similarly, the restarted radio galaxy B2~0258$+$35 
\citep[hosted by NGC 1167][]{shulevski_recurrent_2012,brienza_duty_2018}
has giant lobes $\sim240$~kpc across, and shows evidence
for past/ongoing merger activity \citep{emonts_large-scale_2010,struve_cold_2010}, 
which may have triggered a powerful past jet episode \citep{shulevski_recurrent_2012}.
The offsets from the TA hotspots for NGC~6251 and B2~0258$+$35 are $48.7^\circ$ and $73.8^\circ$, respectively.
Thus, the required magnetic deflections are large, but $48.7^\circ$ is possible for an ${\cal R}\sim10$EV CR  
from a source at NGC~6251's position and distance in either the Galactic magnetic field 
\citep[][see their figure~2]{farrar_galactic_2016} 
or an extragalactic field of $\sim1$nG in accordance with results from \citep{bray_upper_2018}. 
Other sources such as DA 240 and 3C 264 are also interesting candidates.
It is difficult to draw more robust conclusions about arrival directions without detailed modelling that takes into account attenuation losses and magnetic field deflections, which are both highly composition-dependent \citep[e.g.][]{alves_batista_cosmic_2016,wykes_uhecr_2017}; nevertheless, an association of UHECR arrival directions with giant-lobed radio galaxies is at least feasible. 

While a ``dormant source'' scenario was invoked by \cite{matthews_fornax_2018_mnras} to explain UHECR arrival directions,
it is also appealing in terms of source energetics. 
The minimum power requirement (equation~\ref{eq:power}) and minimum source densities (Fig.~\ref{fig:uhecr_lum_density}) provide quite strict limits for UHECR acceleration in steady sources. These limits initially appear problematic when we consider the relative scarcity of nearby (within a GZK radius) FRII/high-power sources \citep[e.g.][]{massaglia_radio_2007,massaglia_role_2007,van_velzen_radio_2012,matthews_fornax_2018_mnras}. As pointed out by, e.g., \cite{nizamov_constraints_2018}, the former constraint is alleviated by allowing for variable sources. In such a situation, the power requirement is no longer instantaneous and instead we require a powerful outburst satisfying $Q>Q_{\mathrm{min}}$ within the shorter of the GZK time ($\sim 300$~Myr) or UHECR escape time. The constraints from Fig.~\ref{fig:uhecr_lum_density} are unchanged, as it is the active source density that matters; we do not necessarily require any sources to be currently active within a characteristic GZK radius, but we require that sources are {\em on average} active enough to produce the observed UHECR flux at Earth. 

The arguments made so far in this section are not specific to exactly where the UHECRs are accelerated in the source, as they only require that there is sufficient power in the jet. If the UHECRs are accelerated in backflows then that imposes an additional limit, in that the physical conditions for backflow must have been met during an outburst phase. Backflows are not confined to FRIIs. The existence of backflows has been inferred in two lobed FRI sources with radio luminosities of $\approx 10^{41}$erg~s$^{-1}$ \citep{laing_relativistic_2012}, while the  FRI-FRII luminosity break is slightly higher, at approximately $4\times10^{41}$erg~s$^{-1}$ \citep{fanaroff_morphology_1974}. Neither of these thresholds should be thought of as a clearly delineated boundary, but they allow us to estimate if sources with strong backflows might be common enough to explain the observed UHECRs at Earth. Comparing the expected UHECR luminosity density from sources above these luminosities to the observed UHECR luminosity density (Fig.~\ref{fig:uhecr_lum_density}) suggests that a hypothesis in which UHECRs originate in FRII or high power, lobed-FRI radio galaxies is plausible. 

\section{Summary \& Conclusions}
We have used hydrodynamic simulations of AGN jets together with some general physical
arguments to show that backflowing streams present in radio galaxy lobes should produce 
moderately strong shocks that can accelerate CRs to ultra-high energies. We 
summarise our work as follows:
\begin{itemize}
\item We have presented 
hydrodynamic simulations showing that strong backflows are expected in the lobes of 
radio galaxies and multiple shocks can occur along the backflow. A Bernoulli-like analysis 
applied to a streamline of steady flow helps elucidate many of the key physical 
effects, showing that the backflow should be thin and supersonic, with its velocity 
a significant fraction ($\sim1/2$) of the jet velocity. 
The Mach number in the backflow increases
as the pressure drops so as to equilibriate with the surrounding pressure in the cocoon. 
Backflows are generaly strongest for high density contrasts, wide lobes and powerful jets and are not confined to our fiducial parameter space.
\item Our 3D simulations show similar overall behaviour to 2D cylindrical simulations in terms of 
backflow strength, but the breaking of azimuthal symmetry allows the kinetic energy to be focused into a stream of smaller cross-sectional area than when assuming a cylindrical geometry.  
\item We have used a combination of Lagrangian (tracer particles) and Eulerian (grid 
properties) techniques to analyse the shocks in our simulations. These methods
reveal characteristic shock sizes of $r_s\sim2$~kpc and shock velocities of $v_s\sim 0.2c$. 
Approximately $10\%$ of tracer particles pass through a shock of $M\gtrsim3$. For a 
magnetic field of $140~\mu G$ this leads to a Hillas energy for protons of $5\times10^{19}$eV 
(or equivalently, maximum rigidities of ${\cal R}\sim50$EV).
\item We have shown that the shocks that form in backflows have a number of key advantages over the relativistic jet termination shock for UHECR acceleration (see section~\ref{sec:magnetic}). The shock velocities inevitably cover a range of values, meaning that many of the problems with DSA at relativistic shocks \citep[e.g.][]{reville_maximum_2014,bell_cosmic-ray_2018} can be avoided. Multiple shocks in the flow also allow for multiple opportunities for acceleration via DSA and magnetic field amplification at the shock. Perhaps even more importantly, multiple shocks along a flow mean that there is more than one way for the magnetic field to be amplified and, crucially, the amplification timescale at the shock is not necessarily the limiting factor for acceleration to high energy. As a result, the Hillas energy can be expected to act as a better estimate of the maximum CR energy in jet backflows than in the case of supernova remnants where the energy is lower than Hillas by a significant factor \citep[e.g.][]{bell_cosmic-ray_2013}.
\item We have used radio galaxy luminosity functions and empirical radio to jet power relationships to show that radio jets are, on average, common and powerful enough to produce the UHECR flux arriving at Earth. However, there are not enough steady powerful sources within a canonical GZK horizon to produce the observed UHECRs. We have therefore expanded on the ideas presented by \cite{matthews_fornax_2018_mnras}, exploring the possibility that UHECRs are produced during powerful past jet episodes and are now slowly escaping from ``dormant'' giant radio sources, such as Fornax A, Centaurus A and NGC 6251. This class of dormant sources may be able to explain the observed anisotropies from the PAO and TA UHECR observatories. 
\end{itemize}
While the scenario we have presented here offers good overall prospects as an UHECR production model, we note that many of the requirements for UHECR acceleration may be met in other situations other than backflows. Fundamentally, regardless of the class of astrophysical system considered, one has to engineer a ``goldilocks'' situation where a large amount of power ($\sim10^{44}$~erg~s$^{-1}$) is channeled through a strong shock meeting the Hillas criterion, without the shock velocity becoming too large. Shocks in the backflows of radio galaxies provide one way to do this.

\section*{Acknowledgements}
We thank the anonymous referee for a helpful report that improved the quality of the paper. 
We would also like to thank A. Watson, L. Morabito, S. 
Kommissarov, R. Hirai, R. Laing, R. Bowler and J. Bray for helpful 
discussions, as well as C. Frohmaier and R. French
for their help with the clustering algorithm. We are grateful to the 
attendees of two conferences -- the plasma astrophysics workshop held 
in Oxford in July 2017 and the computational MHD workshop held in Leeds 
in December 2017 -- for many stimulating talks and conversations.
This work is supported by the Science and Technology Facilities Council 
under consolidated grant ST/N000919/1.
We would like to acknowledge the use of the University of 
Oxford Advanced Research Computing (ARC) facility in carrying out 
this work: http://dx.doi.org/10.5281/zenodo.22558.
We also gratefully acknowledge the use of the following 
software packages: Visit \citep{childs_visit_2005}, 
matplotlib 2.0.0 \citep{matplotlib}, 
PLUTO 4.2 \citep{mignone_pluto:_2007} and
scikit-learn \citep{scikit-learn}.



\bibliographystyle{mnras}


\bsp	
\label{lastpage}
\end{document}